%% file: main.tex
\documentclass[twocolumn]{aastex63}
\usepackage{graphicx}
\usepackage{placeins}
\usepackage{amsmath}
\usepackage{amssymb}
\usepackage{natbib}
\usepackage{xcolor}
\usepackage{rotating}

\newcommand{\um}{\micron\;}

\newcommand{\Msun}{\hbox{M$_\odot$}}
\newcommand{\nh}{$N_{\rm H}$}

\newcommand{\z}{$1<z\lesssim2$\;}

\newcommand{\0}{\phantom{0}}

\newcommand*\cleartoleftpage{%
  \clearpage
  \ifodd\value{page}\hbox{}\newpage\fi
}

\def \az {\cite{Azadi2020}\;}

\def\nar{\ref@jnl{New A Rev.}}          

\shorttitle{The Bolometric Luminosity Correction of Radio-Quiet and Radio-Loud Quasars at $1<z<2$}
\shortauthors{Azadi et al.}

\begin{document}

\title{The Bolometric Luminosity Correction of Radio-Quiet and Radio-Loud Quasars at $1<z<2$}

\author[0000-0001-6004-9728]{Mojegan Azadi}
\affiliation{Center for Astrophysics $|$ Harvard \& Smithsonian, 60 Garden Street, Cambridge, MA, 02138, USA}
\author[0000-0003-1809-2364]{Belinda Wilkes}
\affiliation{Center for Astrophysics $|$ Harvard \& Smithsonian, 60 Garden Street, Cambridge, MA, 02138, USA}
\author[0000-0001-5513-029X]{Joanna Kuraszkiewicz}
\affiliation{Center for Astrophysics $|$ Harvard \& Smithsonian, 60 Garden Street, Cambridge, MA, 02138, USA}
\author[0000-0002-3993-0745]{Matthew Ashby}
\affiliation{Center for Astrophysics $|$ Harvard \& Smithsonian, 60 Garden Street, Cambridge, MA, 02138, USA}
\author[0000-0002-9895-5758]{S. P. Willner}
\affiliation{Center for Astrophysics $|$ Harvard \& Smithsonian, 60 Garden Street, Cambridge, MA, 02138, USA}


\begin{abstract}

To understand the impact of active galactic nuclei (AGN) on their host galaxies and large scale environment it is crucial to determine their total radiative power across all wavelengths (i.e., bolometric luminosity). In this contribution we describe how quasar accretion disk spectral energy distribution (SED) templates, parameterized by the black hole (BH) mass, Eddington ratio, and spin
can be used to estimate their total radiated luminosity. 
To estimate the bolometric luminosity of AGN, we integrate the accretion disk SEDs from 1\,$\mu$m to 10\,keV. Our approach self-consistently covers any gaps in observations and does not include reprocessed emission from the torus. 
The accretion disk SED, and consequently the bolometric correction inferred from it, strongly depend on the BH mass, the  Eddington ratio, and spin. In particular, the bolometric correction in the visible bands (5000\AA\ and 3000\AA) strongly depends on BH mass, and at X-ray strongly depends on the Eddington ratio. At wavelengths closer to the peak of the accretion disk SED the dependence becomes weaker. Additionally, maximally-rotating (spin = 1) quasars require a higher bolometric correction than their non-rotating (spin = 0) counterparts at all wavelengths.
The SEDs and the bolometric correction presented in this work can determine the radiative power of any sample of radio-quiet to radio-loud Type~1 AGN with observations in the range from  1\,\um to 10\,keV provided the observations are corrected for extinction.

\end{abstract}


\input{intro.tex}

\input{data.tex}

\input{bol.tex}

\input{discussion.tex}

\input{summary.tex}

\bibliographystyle{apjurl}
\bibliography{references.bib}
\input{Appendix}
\end{document}

%% file: intro.tex
\section{Introduction} \label{sec:intro}


Active Galactic Nuclei (AGN),  powered by the accretion of gas and dust onto supermassive black holes (SMBHs) at their centers, are among the most powerful objects in the Universe.  AGN present unique observational signatures 
over more than twenty orders of magnitude in wavelength from radio to $\gamma$-ray.     
In practice, observations of individual AGN tend to be restricted to relatively narrow wavelength ranges, affording only limited views of their total radiated power. To understand AGN physics and their impact on their environment (i.e., host galaxy, galaxy cluster), however, it is crucial to determine their total radiative power (i.e. bolometric luminosity) across all wavelengths.

Standard practice when determining AGN bolometric luminosities is to integrate their spectral energy distributions (SEDs) over all observed frequencies. 
SEDs suffer from several well-known limitations. For example, 
observations generally cover limited and discontinuous frequency ranges.  To compensate, studies rely on gap repair \citep[e.g.,][]{Elvis1994,Richards2006}.
Also, although radio emission from AGN (e.g., from associated radio structures such as lobes) is isotropic, mid-infrared (MIR) emission from the torus is less so, and visible--UV emission from accretion disks is strongly anisotropic.
To account for the anisotropic emission, some studies \citep[e.g.,][]{Hubeny2001,R12} consider an average viewing angle and adjust the bolometric luminosity accordingly. However, this simplification is only acceptable in the Newtonian regime.  Relativistic effects such as aberration and beaming introduce more viewing-angle dependence in the observed SED 
\citep[e.g.,][]{Hubeny2001,Nemmen2010,R12}.

There is disagreement on 
the wavelength range that should be considered 
when determining the radiative power of the AGN.  Some studies include both the visible-UV emission from the accretion disk and the MIR emission from the torus \citep[e.g.,][]{Elvis1994,Richards2006}. Others argue that including MIR emission overestimates the bolometric luminosity \citep[e.g.,][]{Marconi2004,Nemmen2010,R12}
by double-counting the visible-UV photons which are scattered and re-emitted in the IR and have already been accounted for in the visible-UV portion of the SED.  

Earlier studies determined the bolometric correction from an average SED \citep[e.g.,][]{Elvis1994}. However, an average SED may not be representative of the general AGN population. Therefore, studies moved towards
generating a range of SEDs around the average based on the correlation between the optical-to-X-ray spectral index $\alpha_{ox}$ and 2500\,\AA\ luminosity \citep{V2003} to construct such accretion disk template \citep[e.g.,][]{Marconi2004,Richards2006,Hopkins2007}.  Some studies found the bolometric correction is luminosity-dependent \citep[e.g.,][]{Marconi2004,Hopkins2007}. In contrast, others 
find instead that the Eddington ratio is a primary parameter impacting the bolometric correction \citep[e.g.,][]{V2007}.

Recently, \az\  introduced a state-of-the-art AGN radio to X-ray SED fitting model (ARXSED) and characterized the SEDs of 
20 radio-loud quasars from the Revised-Third Cambridge Catalogue of Radio Galaxies \citep[3CRR;][]{Laing1983} at \z.  
ARXSED significantly improves on other
SED models \citep[e.g.,][]{Elvis1994,Richards2006,Shang2011,pece15} by including 1) a broader range of photometry spanning radio to X-ray wavelengths, 2) improved AGN component models for the emission from the accretion disk \citep{Kubota2018} and torus \citep{Siebenmorgen2015}, 3) a radio component 
to account for emission from complex radio structures
(lobes, jets, cores, and hot spots) and reproduce a high-energy turnover or cutoff due to the aging of the electron populations, 4) a host galaxy component that replicates galaxy emission from radio to UV wavelengths \citep{dc2008,dcr2015}, and 5) estimates for the \emph{intrinsic} AGN SED obtained by correcting the photometry for reddening and absorption occurring in the torus, the host galaxy, and the Milky Way.
These five improvements offer an opportunity to model AGN emission with new and unprecedented fidelity.

In this contribution we use the SED modeling approach of \az and present the bolometric luminosity correction of both \emph{non-rotating and maximally-rotating AGN}. We first use the \az SED fits (ARXSED) to individual radio-loud quasars (20 quasars) from the 3CRR sample at \z  to determine their bolometric luminosities and correction factors.
ARXSED allows us to overcome the gaps in the photometry, the viewing angle correction, including/excluding MIR data, and thereby accurately estimate the bolometric correction factors. We then expand our analysis to more general samples, building upon the \az approach. We create $\sim$11000 artificial accretion disk templates utilizing QSOSED model of \cite {Kubota2018} by varying BH mass (from $10^7$ to $10^{10}$ $M_{\odot}$), Eddington ratio (from 0.03 to 1.0 in dex) and spin (from 0 to 1) within the acceptable range by the model. We assume a viewing angle of $30^{\circ}$ and $z\sim1.5$. We estimate the bolometric luminosity and bolometric correction for these general quasar population. Finally, we determine quantitatively how the accretion disk parameters (i.e., BH mass, Eddington ratio and spin) impact the shape and normalization of the accretion disk SED and consequently the bolometric correction.

This paper is organized as follows: We describe the study sample and the details of the ARXSED model, and present the bolometric correction calculations of the 3CRR quasars in 
Section~\ref{sec:rl}. We then determine the bolometric correction factor of a general population of AGN from non-spinning to maximally-spinning sources in Section ~\ref{sec:bc_all} and compare our findings with the literature. We summarize our results in Section~\ref{sec:summary}. Throughout the paper we adopt a flat cosmology with $\Omega_{\Lambda}$ = 0.7 and $H_0$= 72 km s$^{-1}$ Mpc$^{-1}$.

%% file: data.tex
\section{Bolometric luminosity corrections for 
$1<z<2$ quasars}
\label{sec:rl}
\subsection{Radio-Loud Sample and Data}

To determine the bolometric correction factor of the most powerful radio-loud quasars at $1<z<2$, we used the quasars from the second revision \citep[3CRR,][]{Laing1983} of the Third Cambridge Catalogue of Radio Galaxies. The full 3CRR catalog includes 173 FR~II radio galaxies up to  $z < 2.5$ and is 96\% complete to a 178\,MHz flux density of 10\,Jy. From $1<z<2$, the 3CRR  includes 38 broad-line and narrow-line radio galaxies \citep{Wilkes2013}. This study uses the 20 broad-line radio galaxies (i.e., quasars) from this sample. 

\az compiled radio to X-ray SEDs of the 20 3CRR \z  quasars by combining new and archival photometry from  multi-frequency radio observation, the SMA, ALMA, {Herschel}, {WISE}, {Spitzer}, {2MASS}, {UKIRT}, {SDSS}, {XMM-Newton}, and {Chandra}.
The resulting complete, randomly oriented sample  has comprehensive multi-wavelength data from radio to X-ray bands, which we employ in the present work. \az fitted the radio-to-X-ray SED of these 20 quasars with the state-of-the-art model of ARXSED, described below.

%% file: bol.tex
\subsection{ARXSED model}
ARXSED is a semi-empirical model that simultaneously replicates the emission from AGN structures including the lobes and jets, torus,  accretion disk, and the host galaxy. At radio wavelengths, the model accounts for the emission from the extended lobes, as well as jets, cores, and hot spots by including either single or double power-laws, a parabola, or a combination of these models. ARXSED accounts for the turn-down in the synchrotron emission by adding an exponential cutoff. At IR wavelengths, ARXSED accounts for the emission from the torus with the \cite{Siebenmorgen2015} two-phase torus model, in which the dust can be distributed in a homogeneous disk, a clumpy medium, or a combination of both.  At visible--UV--X-ray wavelengths, ARXSED uses the recently developed \cite{Kubota2018} model, which treats emission from the accretion disk as originating from three distinct regions. These are 1) an inner region (i.e., corona) with electron temperature $kT_{e} \sim 40$--100\,keV where hard Comptonization occurs and hard X-rays originate; 2) an intermediate region with electron temperature ${\sim}0.1$--1\,keV where soft compotonization occurs and soft X-rays originate, and 3) an outer region where the thermal UV--visible emission originates.  ARXSED accounts for emission from the host galaxy by adding an underlying component from radio to UV wavelengths \citep[MAGPHYS;][]{dc2008, dcr2015}.  For present purposes, ARXSED provides an estimate of the intrinsic SED of each object in our sample after correcting for reddening and absorption occurring in the torus, the host galaxy, and along the line of sight in the Milky Way. 

ARXSED first corrects the photometry at 0.91--13\,\micron\ for Milky Way absorption using the attenuation law $\tau_{\lambda} \propto \lambda^{-0.7} $ from \cite{charlot2000simple}. Then it fits the torus and  corrects the visible--UV photometry for obscuration from the torus. (See equation ~16 of \citeauthor{Azadi2020}, and also see Siebenmorgen et al.\ 2015). 
The X-ray luminosity used in the modeling is the intrinsic luminosity, i.e., corrected for both intrinsic and Galactic absorption \citep{Wilkes2013,Azadi2020}. 
To fit the visible--UV--X-ray SED of the 3CRR quasars, \az used the \cite{Kubota2018} accretion disk model and constructed 11000 templates by varying its primary parameters (SMBH mass, Eddington ratio, spin, and viewing angle).  \citeauthor{Azadi2020} used any available information to set initial parameters (i.e., \ion{Mg}{2} or \ion{C}{4} lines for estimating BH masses). The viewing angle of the accretion disk templates was assumed to be within $\pm 12^{\circ} $ of the best-fit torus model.
ARXSED found a median viewing angle of $52^{\circ}\pm5\arcdeg$ for these quasars.

\begin{figure}[t!]
    \includegraphics[height=\linewidth,angle=90]{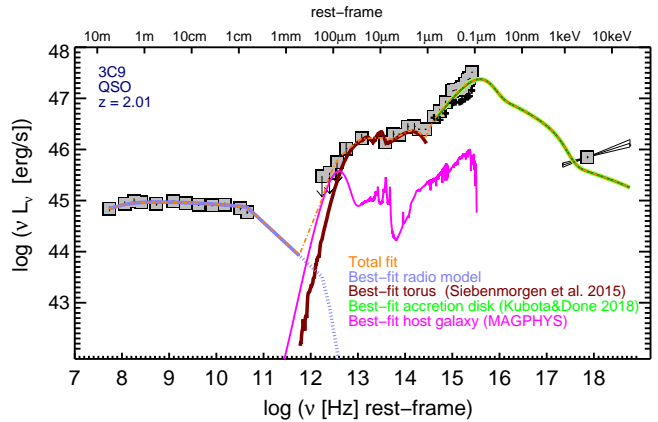}
\caption{The intrinsic SED of 3C~9 from \az. 
Grey boxes show the photometry with black plus signs showing the photometric points before correcting for extinction. Colored lines show contributions to the emission according to the best-fit model.
AGN components are the accretion disk (green), torus (red), and the radio component (light blue).
The radio component includes emission from the lobes, and the dotted curve extending from the radio component traces a cutoff due to aging electron populations.  
Magenta indicates the host galaxy.}    

\label{fig:example}
\end{figure}

\az  presented the best-fitting radio-to X-ray SEDs for 20 quasars at \z from the 3CRR sample (See Table~1 of \citeauthor{Azadi2020}).
These SED fits give estimates of the physical properties of the central engine (SMBH mass, Eddington ratio, spin), torus (viewing angle, dust filling factor, and optical depth), as well as the host galaxy (stellar mass, star formation rate) of all 20 AGN\null.
Figure~\ref{fig:example} presents the radio-to-X-ray SED and ARXSED fit of the quasar 3C~9 in their sample as an example. 

\subsection{Bolometric Corrections for 3CRR Quasars} 
\label{sec:rl_bc}

To estimate the bolometric luminosity and the correction factor as a function of wavelength for each of the 3CRR quasars, we used the best-fit accretion disk model from ARXSED\null. We then integrate over the best-fit accretion disk model, \emph{i.e., the intrinsic rather than the observed SED} \citep{Azadi2020},
from 1\,\um to 10\,keV to determine the bolometric luminosity. The bolometric correction factor is then
\begin{equation} \label{eq:bc}
    BC({\nu}) \equiv \frac{\int _{1\um}^{10keV} L_{\nu} d{\nu}}{\nu L_{\nu}} 
\end{equation}

MIR emission from the torus is not included in the integral because that emission is dominated by reprocessed accretion disk radiation. ARXSED self-consistently covers the gap between the UV and soft X-ray bands without requiring gap-repair techniques. The best-fit accretion disk model thus yields the intrinsic luminosity accounting for the effects of the fitted viewing angle (as well as SMBH mass, Eddington ratio, and spin), and therefore there is no need to apply any further correction due to the anisotropy of the accretion disk observed SED.

\begin{figure}[t!]
\includegraphics[width=0.4\textwidth,angle=90]{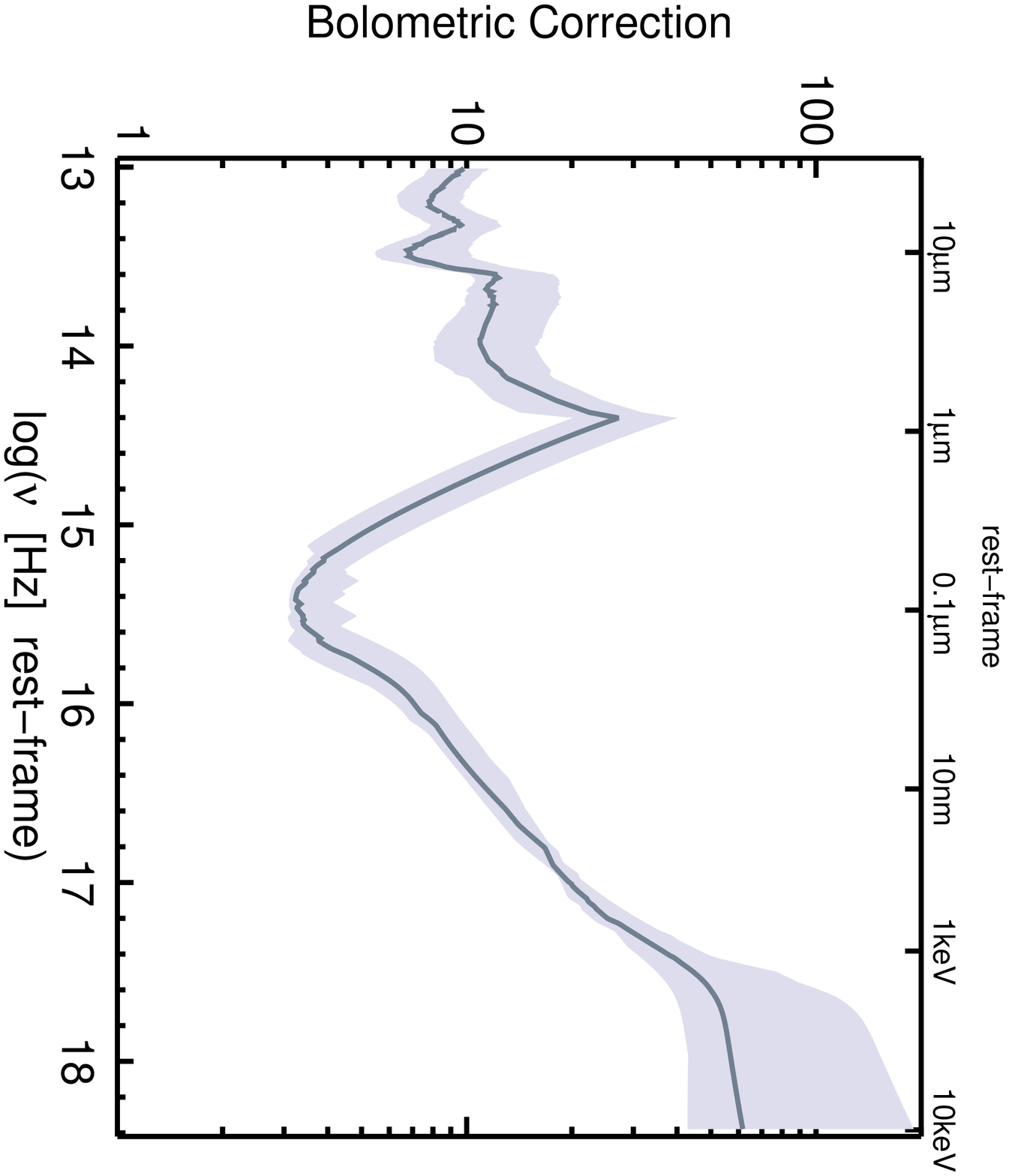}
\caption{Bolometric correction factor as a function of frequency for the 20 3CRR radio-loud quasars at \z \citep{Azadi2020}. The curve shows the median bolometric correction, and the shaded region shows 25th--75th percentile range. 
Correction factors are based on the best-fit models from ARXSED \citep{Azadi2020}: torus at $1\,\micron<\lambda <30$~\micron\ and accretion disk at $\lambda <1$~\micron. 
While the plot extends to  MIR wavelengths, the bolometric luminosity was calculated  only from  1\,\um--10\,keV, i.e., for the intrinsic emission. }    
\label{fig:bc}
\end{figure}

Figure~\ref{fig:bc} presents the bolometric correction as a function of rest-frame frequency for the 3CRR quasars at \z. Although for purposes of illustration the horizontal axis of Figure~\ref{fig:bc} begins at 30\,\micron, to determine the bolometric luminosity we only integrated from 1\,\um  to 10~keV.  At $\lambda >1$\,\micron\ we used the best-fit ARXSED torus model, and at $\lambda <1$\,\micron\ we used the best-fit accretion disk model \citep{Azadi2020}. 
The  bolometric correction has a minimum at $10^{15.4}$\,Hz, which is in the range (10--500\,nm) at which the accretion disk SEDs peak (See Figures~\ref{fig:ad_mass} and~\ref{fig:ad_mdot}).
For observations away from the peak of the accretion disk emission, the bolometric correction is larger, though the uncertainty can remain small.  The sharp peak at rest-frame 1\,\um is due to a dip where the  accretion disk and torus emission both drop. The small dip at rest-frame 9.6\,\um is due to the silicate emission feature in some sources in our sample.

The 25th--75th percentile range in Figure~\ref{fig:bc} increases moving from soft to hard X-ray bands. This is partly because the ARXSED accretion disk fits are not well constrained. The limited number of X-ray data points causes the UV--visible photometry to drive the accretion disk fits.  Additionally, the presence of underlying non-thermal X-ray components related to the jet emission  makes the accretion disk fit more uncertain in the X-ray regime \citep[see][]{Azadi2020}. We will discuss other additional causes for uncertainty in  detail in Section~\ref{sec:bc_general}.


\begin{figure*}[th!]
\begin{center}
\includegraphics[height=0.8\textwidth,angle =90]{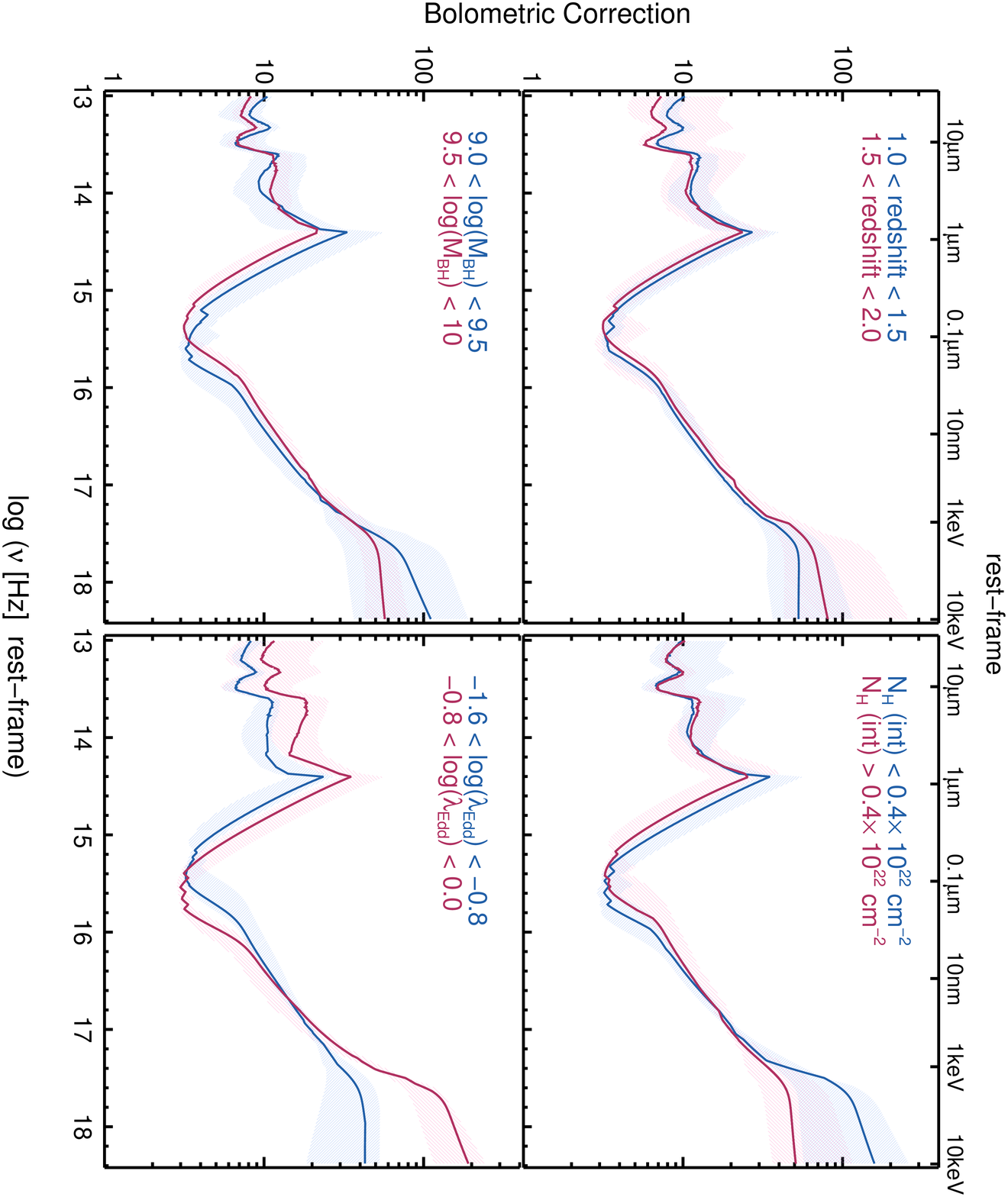}
\caption{Bolometric correction factor as a function of frequency for 3CRR quasars at \z. As labeled, different panels show division by redshift, intrinsic X-ray absorption column density, BH mass, and Eddington ratio.
Given the limited range of spin for these quasars (Section \ref{sec:bc_all}), the sample cannot show any effect of spin. Numerical values are given in Tables~\ref{tab:bc1_3C} and~\ref{tab:bc2_3C}.}    
\label{fig:agn_prop}
\end{center}
\end{figure*} 

\subsection{Bolometric Correction Dependence on  Quasars' Properties}

Figure~\ref{fig:agn_prop} shows the bolometric corrections  for the 3CRR quasars divided into two bins in 
redshift, hydrogen column density \nh, BH mass, and Eddington ratio. The median
bolometric correction in each bin is tabulated in
Tables~\ref{tab:bc1_3C} and~\ref{tab:bc2_3C}. The intrinsic \nh\ values were taken from \cite{Wilkes2013}, and the BH mass and Eddington ratio are from ARXSED fitting (See Table~6 of \citealt{Azadi2020}).
Because of our sample's limited number of sources and limited ranges of parameters, the 25th--75th percentile envelopes overlap at most wavelengths, showing little significant dependence.
The redshift (top-left) panel shows larger uncertainties for the higher-redshift bin around the 9.6\,\um silicate emission feature. This is due to the absence of IRS data  for $z>1.4$ sources \citep{Leipski2010}.  The bolometric correction  minimum frequency is nearly the same in the low- and high-redshift bins, $10^{15.5}$ and $10^{15.4}$~Hz, respectively.

Given that our sources are all quasars, they have a limited range of \nh, and show little differences when they are divided according to their \nh (top-right panel). For sources with low net counts, \nh\ was not well constrained, and a $3\sigma$ upper limit was used instead \citep[see][]{Wilkes2013}. In these cases the intrinsic X-ray luminosity might be over-estimated resulting in lower bolometric correction values at X-ray bands. The bolometric correction has a minimum at $10^{15.6}$ and $10^{15.4}$ Hz in the low and high \nh\  bins respectively.

\begin{figure*}[t!]
\begin{center}
\vspace{0cm}
\includegraphics[width=0.47 \textwidth,angle =90]{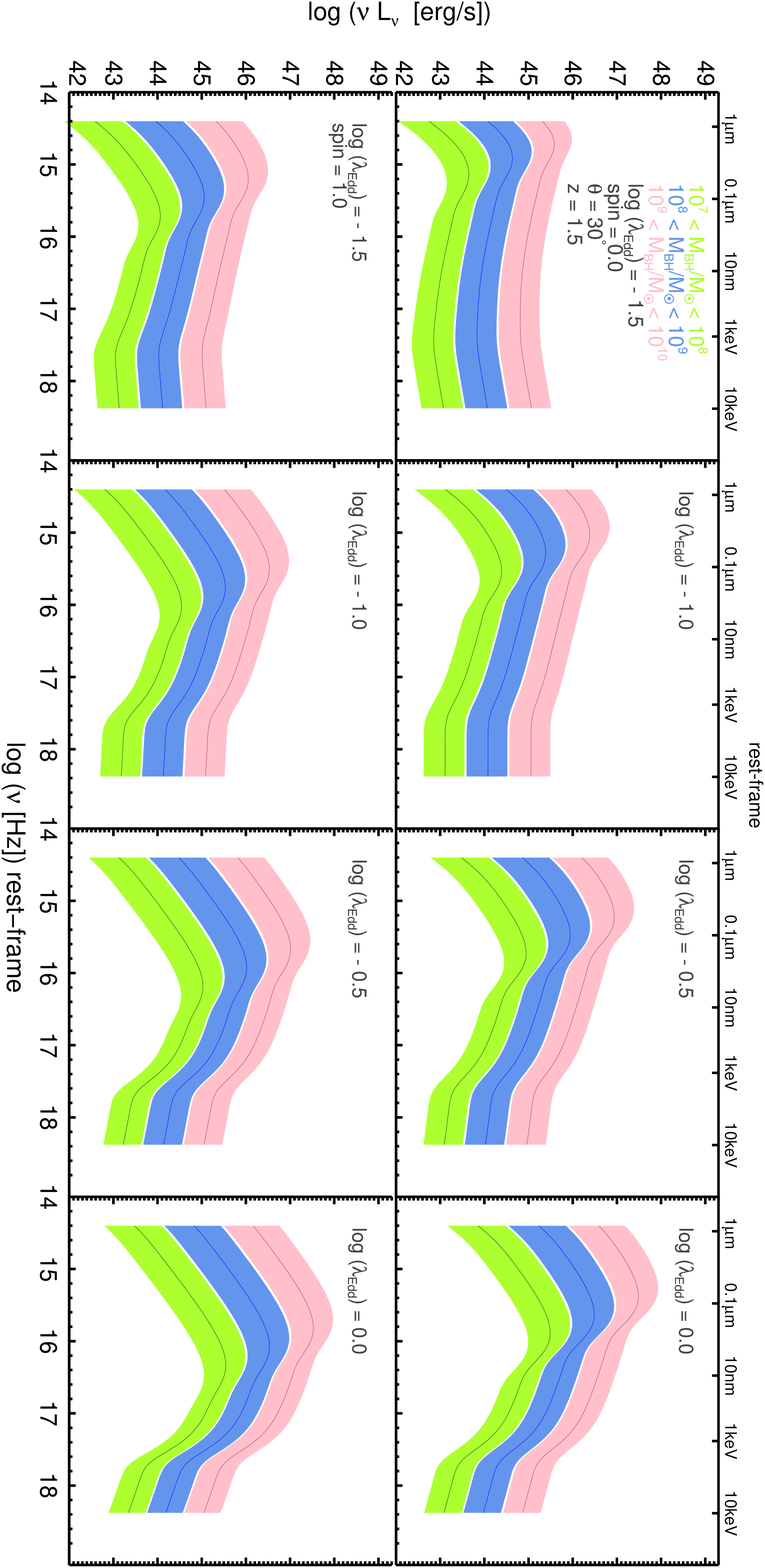} 
\caption{Accretion disk templates created from the \cite{Kubota2018} QSOSED model. In each panel, the BH mass varies from $10^{7}$ to $10^{10}$~\Msun\ with each increasing decade represented by green, blue, and pink shaded areas, respectively.  The Eddington ratio increases from left to right but is fixed in each panel as labeled. Top panels are for a non-rotating BH (spin=0), and  bottom panels are for a maximally-rotating BH (spin=1). The solid lines show the median SED in each mass bin, and the shaded regions show the entire range. An increase in BH mass results in a cooler disk peaking at lower frequencies while an increase in Eddington ratio (and spin) results in a hotter accretion disk peaking at higher frequencies. Numerical values are given in Table~\ref{tab:fig5}.} 
\label{fig:ad_mass}
\vspace{0.5cm}
\includegraphics[width=0.47 \textwidth,angle =90]{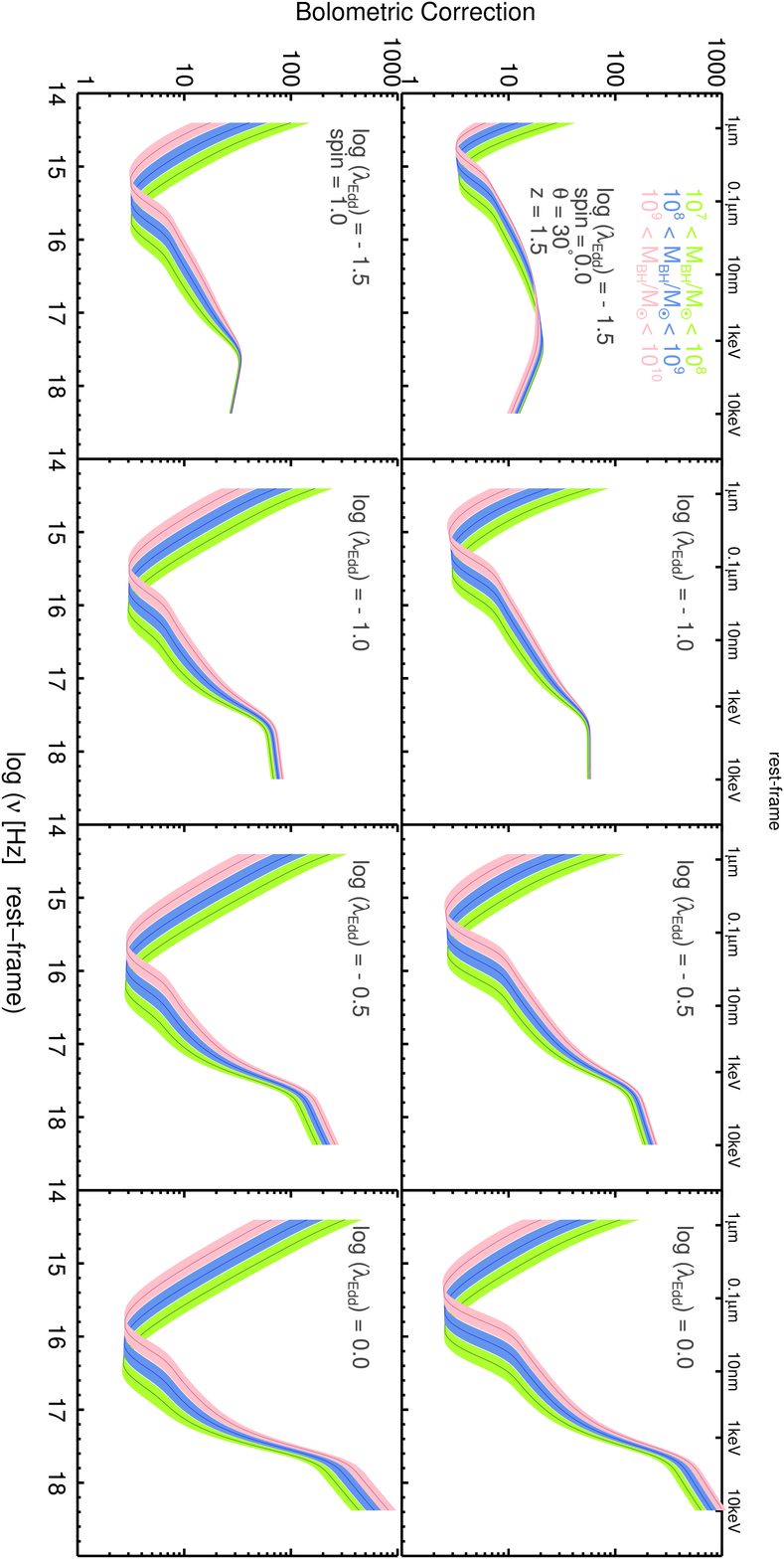}
\caption{Bolometric correction factors as a function of frequency. Bolometric luminosities are calculated by integrating the templates in Figure~\ref{fig:ad_mass} from 1~\um to 10~keV. Green, blue, and pink shaded areas show increasing decades of BH mass as shown in the legend.
The solid lines show the bolometric correction in each mass bin, and the shaded regions show the entire range. The Eddington ratio increases from left to right but is fixed in each panel as labeled. Top panels are for a non-rotating BH (spin=0), and bottom panels are for a maximally-rotating BH (spin=1). Numerical values are given in Table~\ref{tab:fig6}.}    
\label{fig:bc_mass}
\end{center} 
\end{figure*}

\begin{figure*}[th!]
\begin{center}
\vspace{0cm}
\includegraphics[width=0.47\textwidth,angle =90]{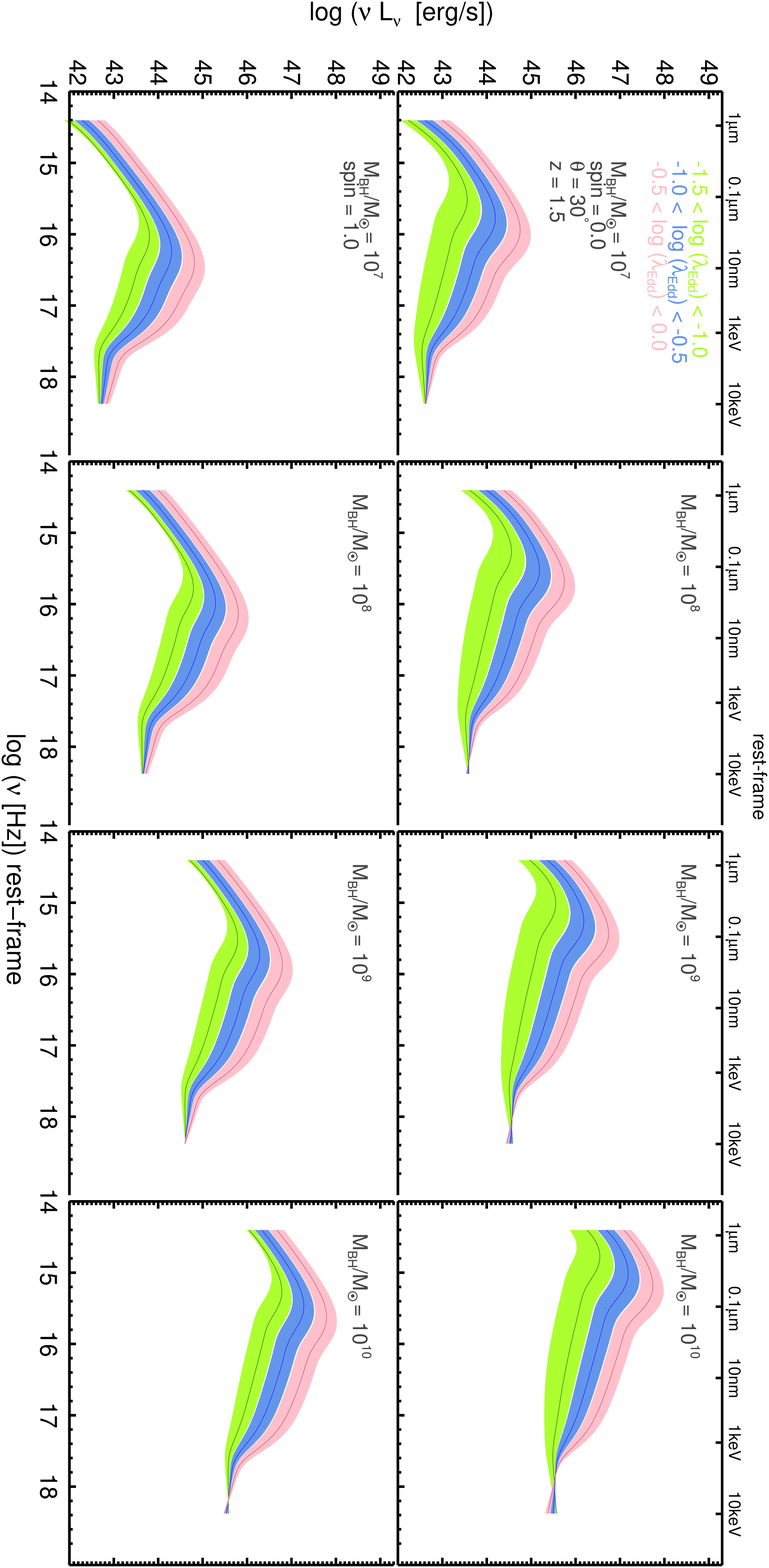} 
\caption{The accretion disk templates created from the \cite{Kubota2018} QSOSED model. $\log(\lambda_{\rm Edd})$ varies from $-1.5$ to 0.0 with green, blue and pink shaded areas showing increasing decades. The BH mass is fixed in each panel and increases from left to right. Top panels are for a non-rotating BH (spin=0), and bottom panels are for a maximally-rotating BH (spin=1). Solid lines show the median SED in each mass bin, and the shaded regions show the entire range. An increase in Eddington ratio (and spin) results in a hotter accretion disk peaking at higher frequencies, while an increase in BH mass results in a cooler disk peaking at lower frequencies. Numerical values are given in Table~\ref{tab:fig7}.} 
\label{fig:ad_mdot}
\vspace{0.5cm}  
\includegraphics[width=0.47\textwidth,angle =90]{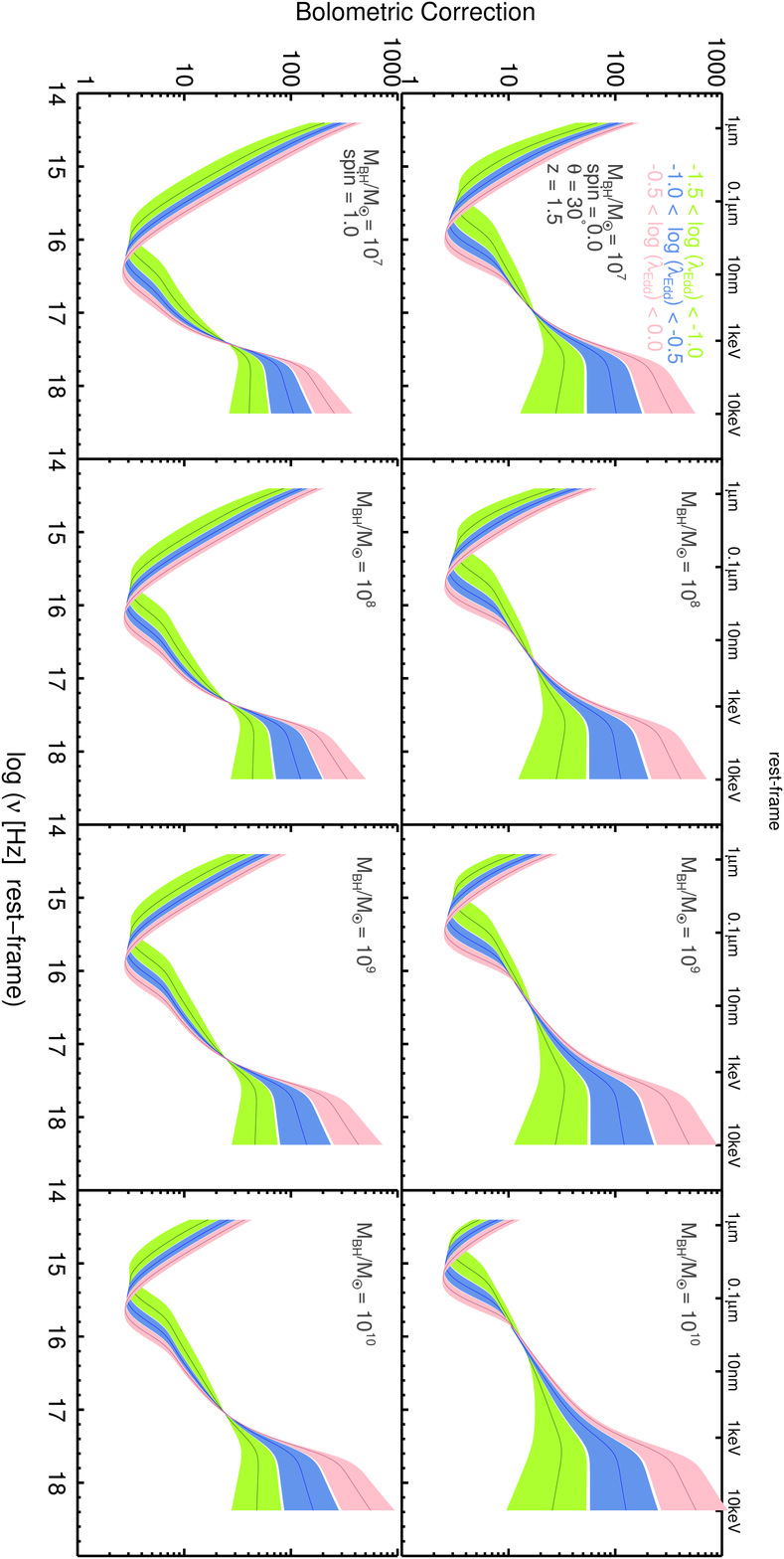}
\caption{Bolometric correction factors as a function of frequency.  Bolometric luminosities are
calculated by integrating the templates in Figure~\ref{fig:ad_mdot} from 1~\micron\ to 10~keV. Green, blue, and pink shaded areas show increasing half-decades of $\log(\lambda_{\rm Edd})$ as indicated in the legend. 
The solid lines show the bolometric correction in each Eddington ratio bin, and the shaded regions show the entire range. BH mass increases from left to right but is fixed in each panel as labeled. Top panels are for a non-rotating BH (spin=0), and bottom panels are for a maximally-rotating BH (spin=1). Numerical values are given in Table~\ref{tab:fig8}.} 
\label{fig:bc_mdot}
\end{center}
\end{figure*}

The small differences related to SMBH masses and Eddington ratios (bottom panels of Figure~\ref{fig:agn_prop}) can be understood with the help of Figures~\ref{fig:ad_mass} and~\ref{fig:ad_mdot}, 
which show how varying BH mass, Eddington ratio, and spin impacts the SEDs. As BH mass increases, the accretion disk emission peak moves towards lower frequencies. This is seen in the bolometric correction minimum being at $10^{15.4}$\,Hz for the higher-mass bin but  $10^{15.6}$\,Hz for the low-mass bin. Eddington ratio has a more noticeable effect. The minimum bolometric correction occurs at  $10^{15.4}$ and $10^{15.5}$~Hz, respectively, in the low- and high-Eddington-ratio bins. This is because
higher Eddington ratio results in a more luminous accretion disk peaking at higher energies (Figures~\ref{fig:ad_mass} and~\ref{fig:ad_mdot}). 

\subsection{Advantages and Limitations of Determining the Bolometric Correction from ARXSED}
\label{sec:lim}

Obtaining bolometric luminosities from ARXSED has a number of advantages.  For example, there is often a large gap in data coverage between the far-UV and X-ray because neither is observable from the ground.  Most studies bridge this gap by interpolating in $\log\nu$--$\log(\nu L_{\nu})$ space \citep[e.g.,][]{Elvis1994,Richards2006,R12}. 
ARXSED self-consistently models the entire radio to X-ray range,  predicting the shape of the SED in any photometric gap \citep[see the median SED in][]{Azadi2020}.

Powerful AGN are assumed to have optically thick, geometrically thin accretion discs \citep[e.g.,][]{SS1976, NT1973} that radiate strongly but not isotropically from
visible--UV to X-ray wavelengths.  In calculating the radiative power of an AGN, this anisotropy should be taken into account.  In a Newtonian regime, the observed flux density is a function of the viewing angle, $\theta$ ($F_{\nu} \propto \cos(\theta)$, 
\citealt{FRK2002}). Most studies correct their observed SED assuming an average viewing angle in a Newtonian regime \citep[e.g.,][]{R12}. 
However, this is an oversimplification, and relativistic effects such as beaming, aberration, and light bending may result in a more complicated observed spectrum and significant departures from the simple Newtonian assumptions \citep[e.g.,][]{Hubeny2001,Nemmen2010}. In fact, \cite{Nemmen2010} found that when relativistic effects are taken into account, the bolometric luminosity can vary from 67\% to 200\% of the integrated isotropic luminosity. 

The ARXSED accretion disk model \citep[QSOSED from][]{Kubota2018} does
not include relativistic effects such as strong reflection and strong
relativistic smearing to
explain the ``soft-X-ray excess" in AGN observed spectra, and assumes that the accretion disk truncates at regions where the hard X-ray emission originates \citep[e.g.,][]{Yaqoob2016,Porquet2018}. 
Instead the model includes a
warm comptonization region to explain the soft-X-ray
excess which seems to be favored over the reflection model by recent
results \citep[e.g.,][]{Porquet2018}. Because the ARXSED accretion disk templates are built based on the viewing angle of our sources (within $\pm12^{\circ}$ of the torus), no further correction for the anisotropy of the accretion disk emission is required in our analysis.

\begin{table*}
\centering
    \caption{Bolometric correction factors for 3CRR quasars at \z (see Figure \ref{fig:agn_prop}).}
    \begin{tabular}{ccccc}
    \hline \hline
        $\rm log (\nu)$  rest-frame &$1.0<z<1.5$ & $1.5<z<2.0$&$N_{\rm H}<0.4\times10^{22}$\,cm$^{-2}$ &$N_{\rm H}>0.4\times10^{22}$\,cm$^{-2}$ 
        \\
         \hline 
        \vspace{0.01pt}
$13.00$&$10.3^{11.1}_{8.23}$ & $7.30^{19.1}_{4.82}$ & $9.82^{19.1}_{7.30}$ & $10.3^{11.1}_{7.99} $\vspace{1pt}\\
$13.25$&$8.68^{10.6}_{7.58}$ & $6.95^{18.5}_{5.41}$ & $8.68^{18.5}_{6.71}$ & $8.62^{10.6}_{7.58} $\vspace{1pt}\\
$13.50$&$6.85^{8.79}_{6.54}$ & $5.80^{14.2}_{5.30}$ & $6.91^{14.2}_{5.52}$ & $6.85^{10.4}_{5.80} $\vspace{1pt}\\
$13.75$&$12.3^{16.4}_{10.4}$ & $11.4^{21.3}_{8.29}$ & $11.6^{21.3}_{9.91}$ & $12.3^{18.3}_{10.4} $\vspace{1pt}\\
$14.00$&$11.1^{14.8}_{8.02}$ & $10.5^{15.6}_{7.22}$ & $11.0^{15.6}_{9.77}$ & $11.1^{16.2}_{8.02} $\vspace{1pt}\\
$14.25$&$18.0^{23.4}_{11.9}$ & $16.4^{24.4}_{10.9}$ & $18.0^{24.4}_{16.4}$ & $18.3^{35.3}_{11.2} $\vspace{1pt}\\
$14.50$&$20.0^{29.0}_{15.0}$ & $17.1^{25.8}_{7.66}$ & $25.8^{41.4}_{22.0}$ & $18.7^{22.7}_{10.6} $\vspace{1pt}\\
$14.75$&$10.1^{14.2}_{7.68}$ & $8.58^{12.8}_{3.96}$ & $12.8^{20.1}_{10.7}$ & $9.28^{11.1}_{5.35} $\vspace{1pt}\\
$15.00$&$5.53^{7.43}_{4.29}$ & $4.79^{6.67}_{3.10}$ & $6.67^{10.2}_{5.67}$ & $5.10^{5.86}_{3.59} $\vspace{1pt}\\
$15.25$&$3.63^{4.46}_{3.47}$ & $3.49^{4.43}_{3.20}$ & $3.93^{5.62}_{3.49}$ & $3.62^{4.43}_{3.41} $\vspace{1pt}\\
$15.50$&$3.29^{3.70}_{3.12}$ & $3.30^{4.98}_{2.79}$ & $3.30^{4.75}_{3.07}$ & $3.37^{5.71}_{3.16} $\vspace{1pt}\\
$15.75$&$4.58^{4.84}_{3.60}$ & $3.55^{5.35}_{3.22}$ & $3.55^{5.35}_{3.09}$ & $4.75^{8.02}_{3.95} $\vspace{1pt}\\
$16.00$&$7.07^{8.50}_{6.22}$ & $6.72^{7.77}_{5.16}$ & $6.54^{7.77}_{4.43}$ & $7.40^{9.67}_{6.68} $\vspace{1pt}\\
$16.25$&$8.75^{11.0}_{7.59}$ & $9.35^{9.63}_{6.57}$ & $8.58^{9.63}_{7.45}$ & $9.09^{11.7}_{8.51} $\vspace{1pt}\\
$16.50$&$11.1^{13.7}_{9.87}$ & $11.8^{12.3}_{8.60}$ & $11.3^{12.3}_{9.87}$ & $11.7^{14.5}_{10.6} $\vspace{1pt}\\
$16.75$&$14.5^{16.2}_{13.4}$ & $15.4^{16.5}_{11.5}$ & $15.4^{16.5}_{13.4}$ & $15.1^{17.6}_{13.5} $\vspace{1pt}\\
$17.00$&$19.5^{19.9}_{17.7}$ & $21.0^{22.8}_{16.1}$ & $20.6^{22.8}_{19.2}$ & $19.6^{21.3}_{17.7} $\vspace{1pt}\\
$17.25$&$27.3^{28.9}_{24.5}$ & $27.1^{38.9}_{25.2}$ & $29.7^{38.9}_{26.0}$ & $27.1^{28.9}_{23.5} $\vspace{1pt}\\
$17.50$&$44.0^{61.1}_{32.1}$ & $45.1^{76.6}_{38.0}$ & $76.6^{76.6}_{44.4}$ & $39.6^{61.1}_{27.8} $\vspace{1pt}\\
$17.75$&$53.0^{85.2}_{33.9}$ & $63.6^{159.}_{43.0}$ & $115.^{159.}_{53.0}$ & $46.7^{85.2}_{32.2} $\vspace{1pt}\\
$18.00$&$53.1^{97.0}_{33.9}$ & $72.3^{196.}_{43.0}$ & $132.^{196.}_{54.3}$ & $48.4^{97.0}_{33.9} $\vspace{1pt}\\
$18.25$&$53.0^{109.}_{35.6}$ & $81.4^{236.}_{42.9}$ & $148.^{236.}_{56.1}$ & $49.9^{109.}_{35.6} $\vspace{1pt}\\
$18.50$&$53.0^{116.}_{36.5}$ & $86.9^{261.}_{42.9}$ & $158.^{261.}_{57.1}$ & $50.9^{116.}_{36.5} $\vspace{1pt}\\
\hline
    \end{tabular}
    \label{tab:bc1_3C}
\end{table*}

\begin{table*}
    \centering
        \caption{Bolometric correction factors for 3CRR radio-loud quasars at \z (see Figure \ref{fig:agn_prop}).} 
    \begin{tabular}{ccccc}
    \hline \hline
        $\rm log (\nu)$  rest-frame &$9.0<\log(M_{\rm BH})<9.5$&$9.5<\log(M_{\rm BH})<10$ &$-1.6<\log(\lambda_{\rm Edd})<-0.8$&$-0.8<\log(\lambda_{\rm Edd})<0.0$  
        \\
         \hline 
$13.00$&$10.5^{11.6}_{7.16}$ & $8.29^{10.3}_{7.80}$ & $8.29^{10.3}_{7.80}$ & $11.6^{24.1}_{7.16} $\vspace{1pt}\\
$13.25$&$9.21^{10.9}_{6.39}$ & $8.02^{8.62}_{6.95}$ & $8.02^{8.68}_{6.95}$ & $10.9^{20.6}_{6.50} $\vspace{1pt}\\
$13.50$&$6.75^{10.4}_{5.30}$ & $6.85^{8.79}_{5.80}$ & $6.75^{6.91}_{5.52}$ & $10.4^{14.9}_{5.43} $\vspace{1pt}\\
$13.75$&$10.4^{18.3}_{7.59}$ & $11.9^{15.2}_{11.4}$ & $10.7^{12.3}_{8.29}$ & $18.3^{23.9}_{11.4} $\vspace{1pt}\\
$14.00$&$9.77^{16.2}_{6.33}$ & $11.0^{14.8}_{10.7}$ & $10.5^{11.1}_{7.22}$ & $15.6^{21.6}_{11.0} $\vspace{1pt}\\
$14.25$&$18.6^{24.4}_{9.50}$ & $16.4^{18.3}_{15.5}$ & $11.9^{18.0}_{9.50}$ & $23.4^{32.0}_{18.3} $\vspace{1pt}\\
$14.50$&$23.7^{41.4}_{18.7}$ & $15.9^{19.1}_{10.7}$ & $17.1^{20.0}_{10.6}$ & $25.8^{41.4}_{22.0} $\vspace{1pt}\\
$14.75$&$11.4^{20.1}_{9.28}$ & $8.05^{9.65}_{5.62}$ & $8.58^{10.1}_{5.35}$ & $12.8^{20.1}_{10.7} $\vspace{1pt}\\
$15.00$&$6.02^{10.2}_{5.10}$ & $4.54^{5.28}_{3.59}$ & $4.79^{5.53}_{3.59}$ & $6.67^{10.2}_{5.67} $\vspace{1pt}\\
$15.25$&$4.43^{5.62}_{3.60}$ & $3.32^{3.63}_{3.20}$ & $3.62^{4.30}_{3.32}$ & $3.93^{5.62}_{3.49} $\vspace{1pt}\\
$15.50$&$3.41^{4.98}_{3.07}$ & $3.29^{3.37}_{2.85}$ & $3.37^{5.71}_{3.21}$ & $2.86^{3.57}_{2.85} $\vspace{1pt}\\
$15.75$&$3.60^{4.84}_{3.01}$ & $4.75^{5.35}_{4.58}$ & $5.30^{8.02}_{4.58}$ & $3.22^{3.55}_{3.01} $\vspace{1pt}\\
$16.00$&$6.54^{7.40}_{4.04}$ & $7.58^{8.50}_{7.07}$ & $7.58^{9.67}_{7.07}$ & $5.16^{6.68}_{4.04} $\vspace{1pt}\\
$16.25$&$8.39^{9.38}_{6.44}$ & $9.35^{11.0}_{8.58}$ & $9.35^{11.7}_{8.51}$ & $7.45^{9.09}_{6.44} $\vspace{1pt}\\
$16.50$&$10.7^{12.2}_{8.19}$ & $11.8^{13.7}_{11.1}$ & $11.8^{13.9}_{10.7}$ & $9.87^{11.7}_{8.19} $\vspace{1pt}\\
$16.75$&$14.0^{16.2}_{11.0}$ & $15.4^{16.7}_{14.5}$ & $15.1^{16.7}_{14.0}$ & $13.4^{15.5}_{11.0} $\vspace{1pt}\\
$17.00$&$19.2^{21.3}_{15.5}$ & $19.9^{21.0}_{19.6}$ & $19.6^{20.6}_{18.6}$ & $19.8^{22.8}_{16.1} $\vspace{1pt}\\
$17.25$&$27.3^{33.9}_{23.5}$ & $27.1^{29.7}_{25.2}$ & $26.0^{28.3}_{22.3}$ & $33.9^{38.9}_{26.0} $\vspace{1pt}\\
$17.50$&$48.6^{76.6}_{27.8}$ & $44.0^{45.1}_{34.9}$ & $34.9^{44.0}_{24.6}$ & $76.6^{76.6}_{61.1} $\vspace{1pt}\\
$17.75$&$74.3^{133.}_{32.2}$ & $52.3^{63.6}_{41.2}$ & $41.2^{52.3}_{22.9}$ & $133.^{143.}_{85.2} $\vspace{1pt}\\
$18.00$&$87.7^{155.}_{33.9}$ & $54.3^{72.3}_{43.0}$ & $43.0^{53.1}_{20.9}$ & $155.^{179.}_{97.0} $\vspace{1pt}\\
$18.25$&$102.^{177.}_{35.6}$ & $56.1^{81.4}_{42.9}$ & $42.9^{53.0}_{19.2}$ & $177.^{220.}_{109.} $\vspace{1pt}\\
$18.50$&$111.^{191.}_{36.5}$ & $57.1^{86.9}_{42.9}$ & $42.9^{53.0}_{18.4}$ & $191.^{246.}_{116.} $\vspace{1pt}\\

\hline
    \end{tabular}
    \label{tab:bc2_3C}
\end{table*}

\begin{figure}[t!]
\includegraphics[width=0.42\textwidth,
angle=90]{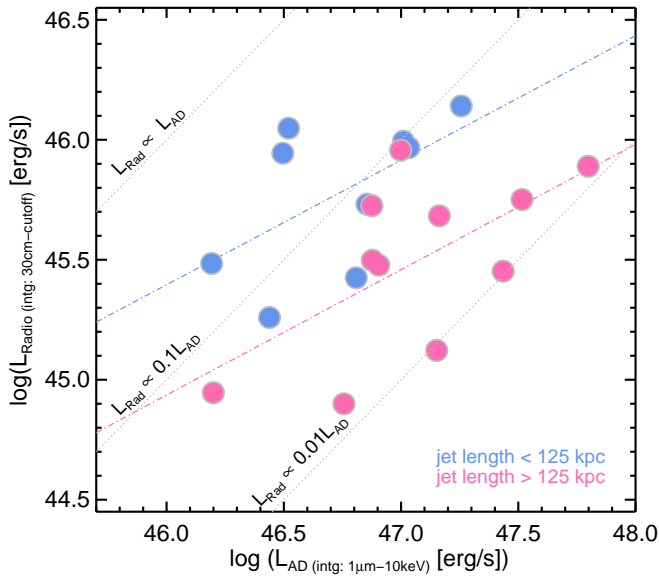}
\caption{The integrated radio luminosity vs.\  integrated accretion disk luminosity in younger, less extended (blue) and mature, more extended (magenta) quasars. 
The dashed-dotted lines show the best fit to each set of  AGN (Equations~\ref{eq:blue} and~\ref{eq:pink}). 
Upper limits are indicated  for sources with uncertain de-projected jet length. For two sources (3C~287/325) with unknown inclination angles,  projected length was used as a lower limit for jet length. (See Table~6 of \citealt{Azadi2020}.)}
\label{fig:rad_vs_ad}
\end{figure}

As noted in Section~\ref{sec:intro}, there is a disagreement in the literature over the range of photometry that determines AGN  bolometric luminosity. Some studies suggest that integrating over the visible-to-soft X-ray range provides a reliable estimate  \citep[e.g.,][]{Nemmen2010, R12}. \cite{R12} argued that under the assumption of isotropy, no re-processed emission should be included in the bolometric luminosity calculation. MIR photons from the torus are nearly all reprocessed visible--UV photons from the accretion disk, which are already accounted for. By contrast, \cite{Richards2006} included the MIR--X-ray emission; they argued that the integrated MIR emission may be a better indicator of bolometric luminosity given that MIR emission is more isotropic than the accretion disk emission. One of the advantages of our approach is that ARXSED first fits the torus templates and corrects the visible--UV photometry for obscuration based on that.  Therefore, our integration (1\,\um--10\,keV) already includes all the seed photons that lead to the MIR emission of the torus. 

Despite its advantages, ARXSED has some limitations. The main limitation is a lack of reliable photometry in some bands.  Several sources in our sample (e.g., 3C\,325, 4C\,16.49) lack reliable UV--visible photometry and/or have very few data points, leaving their accretion disk parameters poorly constrained. 
Additionally, the accretion disk model of \cite{Kubota2018} underestimates the hard X-ray emission in most of our sources. While one possible explanation is the presence of strong, radio-core-related, inverse Compton emission at higher energies \citep{Wilkes1987}, another possibility is that the \cite{Kubota2018} accretion disk model underestimates the coronal X-ray emission of radio-loud sources \citep{Azadi2020}. Given that this particularly occurs in sources with strong radio core emission, the former is more likely. This underlying component leaves ARXSED fits relatively poorly constrained in the hard X-ray regime.

\subsection{The Integrated Power at Radio Frequencies}

Our sample of 3CRR quasars at \z\ are among the most luminous radio sources at this epoch.
Figure~\ref{fig:rad_vs_ad} shows the dependence of the total radio luminosity on the accretion disk luminosity in our sample. The radio luminosities were calculated from the best-fit radio model, integrated from $10^{8}$\,Hz to the cutoff frequency (Table~7 of \citealt{Azadi2020} gives model parameters). The cutoff frequency
occurs where energy losses in the electron population result in a rapid decrease in radio emission 
\citep[e.g.,][]{Blandford1979, Konigl1981}. This cutoff frequency in our sources occurs between $10^{11}$ and $10^{13}$\,Hz and is higher in sources with strong core emission \citep{Azadi2020}. The integrated radio luminosity includes contributions from all components, i.e., the lobes, jets, and cores.

In Figure~\ref{fig:rad_vs_ad}, we divide our sample into two groups of similar size based on the de-projected jet lengths (to those with de-projected length  $<125$\,kpc and $>125$\,kpc, \citealt{Azadi2020}) to show the relationship between the integrated radio and accretion disk luminosity for quasars with short jets (plotted in blue) and quasars with long jets (plotted in pink).
Radio luminosity depends somewhat on jet length, as indicated in Figure~\ref{fig:rad_vs_ad}. In general, quasars with less extended jets are thought to be younger or have jets frustrated by large amounts of surrounding material.  These objects tend to show higher radio emission relative to their accretion disk luminosity than quasars with more extended (i.e., older) jets. The best linear fits to the compact/young and extended/mature sources in our sample are given in equations~\ref{eq:blue} and \ref{eq:pink} respectively:

\vspace{-0.5cm}
\begin{equation}
\log (L_{\rm Radio})=
     (0.52\pm0.29) \log (L_{\rm AD}) - ((21.5\pm13.4)
 \label{eq:blue}
\end{equation}
\vspace{-0.6cm}
\begin{equation}
  \log (L_{\rm Radio})=  (0.52\pm0.22) \log (L_{\rm AD}) -(20.9\pm10.5)
\label{eq:pink}
\end{equation}

For most sources, the radio power is between~1\%--10\% of the accretion disk power, but in three outlying sources with shorter jets (3C\,43/287/318), the radio power is $\sim$30\% of the accretion disk power. A closer look at the 
SEDs of these three sources reveals a relatively flat, UV--visible--IR SED \citep{Azadi2020}, possibly due to an additional non-thermal component from the jet that is not accounted for by ARXSED. The presence of an additional component may 
impact the best-fit torus and accretion disk fit and the estimated luminosity. 

Figure~\ref{fig:rad_vs_ad} shows that the radio size and accretion disk luminosity are linked with sources having $\log (L_{\rm AD})<46.9$ mainly being compact (blue).
These sources also show more radio-related X-ray emission, which includes synchrotron self-Compton (SSC) and inverse Compton (IC) emission.
These components are not accounted in the ARXSED fits but were estimated  
by subtracting the best-fit accretion disk model from the X-ray data.
This may suggest that the more compact radio sources have intrinsically higher IC and/or SSC X-ray emission.  
If we include the radio-linked X-ray emission, the total radiative power from the radio structures would be higher, increasing the observed difference in Figure~\ref{fig:rad_vs_ad} between the large and small radio sources.

Our sample is by construction biased against sources with low radio and accretion disk luminosity.  However, in Figure~\ref{fig:rad_vs_ad}, on average, the smaller sources have higher radio luminosity than the extended ones. This suggests that compact radio sources may be intrinsically more luminous radio emitters than extended sources, and as the radio structures expand and age, their radio luminosities decline.

In summary, both small and extended radio sources in our sample show radio luminosity proportional to the square root of accretion disk luminosity (Equations~\ref{eq:blue} and~\ref{eq:pink}). Neither of the two subsets shows a redshift dependence. However, the SSC/IC emission originating from the radio structures may affect the slopes in each population differently, thus changing the slopes. Additionally, the integrated radio luminosity includes emission from different structures (i.e., lobes, jets, cores, hot spots) and therefore includes radio emission from different epochs of the AGN's history.  In contrast, the accretion disk luminosity is presumably related to the current mass accretion. Therefore the two axes of 
Figure~\ref{fig:rad_vs_ad} compare the radiative powers on different timescales. While our findings suggest that small radio sources are brighter radio-emitters than their extended counterparts, understanding what triggers the relations in Figure~\ref{fig:rad_vs_ad} merits further analysis with larger samples.

%% file: discussion.tex
\section{Bolometric Luminosity Corrections for Radio-Loud and Radio-Quiet Quasars} \label{sec:bc_all} 

This section describes the accretion disk SED and the bolometric correction factors for a general population of quasars (with spin 0 or 1) at $z\sim$1.5, and then compares those bolometric corrections to those found in the literature.

\subsection{Bolometric Correction Factor for a General Quasar Population at $z\sim1.5$} \label{sec:bc_general}

To extend the analysis beyond the limited 3CRR sample, we use the QSOSED model \citep{Kubota2018} to create $\sim$11000 templates corresponding to BH masses ranging from $10^7$ to $10^{10}$~$M_{\odot}$ in steps of 0.05~dex, Eddington ratio from  0.03 to 1.0 in steps of 0.025 dex, and spin from 0.0 to 1.0 in steps of 0.5.\footnote{In the QSOSED model, the range for the BH mass is $10^7$ to $10^{10}$~$M_{\odot}$, for $\log\lambda_{\rm Edd}$ is $-$1.65 to 0.39, and for spin is 0 to~1.} We assumed a $30^{\circ}$ viewing angle appropriate for quasars, which are viewed relatively face-on, and limited our analysis to $z\sim1.5$. Rather than representing a specific quasar sample, these artificial templates replicate the accretion disk emission of a general AGN population at $z\sim1.5$.
The results are presented in Figures~\ref{fig:ad_mass} -- \ref{fig:bc_mdot}  and tabulated in Tables \ref{tab:fig5}--\ref{tab:fig8} in the Appendix.

Figures~\ref{fig:ad_mass} and~\ref{fig:ad_mdot} show how the QSOSED model accretion disk SEDs
respond to variations in the SMBH mass and Eddington ratio, respectively, holding the
other parameters fixed. An increasing SMBH mass results in a cooler and more luminous accretion disk (Figure~\ref{fig:ad_mass}). Increasing the 
Eddington ratio (Figure~\ref{fig:ad_mdot}) or spin (lower panels of Figures~\ref{fig:ad_mass} and~\ref{fig:ad_mdot}) results in a more luminous and hotter disk (Section 3.1 of \citealt{Azadi2020} provides a more detailed discussion).

Figures~\ref{fig:bc_mass} and~\ref{fig:bc_mdot}
show the bolometric corrections corresponding to the above parameter sets as a function of frequency. Given the large ranges of the bolometric correction in some panels, they are plotted on a logarithmic scale. The bolometric luminosities are calculated by integrating the accretion disk templates of Figures~\ref{fig:ad_mass} and \ref{fig:ad_mdot}, respectively, from 1\,\um to 10\,keV as explained in Section~\ref{sec:rl_bc}.   The effects of varying BH mass, Eddington ratio, and spin on the accretion disk SED and the bolometric correction are inter-related. Obtaining independent estimates of one or more of these parameters will result in more accurate bolometric correction estimates.

Figure~\ref{fig:bc_mass} 
shows that the bolometric correction  has a minimum
at $10^{15}$--$10^{16}$\,Hz, corresponding to the peak of the accretion disk SED. As the BH mass increases, 
this minimum shifts to lower frequencies because more massive BHs
produce cooler accretion disks (Figure~\ref{fig:ad_mass}).  
At frequencies below the minimum, more massive BHs require a smaller bolometric correction than their 
lower-mass counterparts. This is because the lower frequencies are closer to the peak of the accretion disk emission for more massive BHs. 
At frequencies above the minimum, the opposite happens: smaller bolometric corrections are needed for lower BH masses, because the peak of accretion disk emission occurs at higher frequencies. 
BH mass has the least effect on the bolometric correction near the minimum value and also at the X-ray frequencies, where the correction is strongly dependent on Eddington ratio but weakly dependent on BH mass. In the 1--10\,keV range, the integrated (bolometric) luminosity increases as Eddington ratio increases (from left to right in the Figure \ref{fig:bc_mass}) so the differences among the three colors, corresponding to different mass ranges, become more distinct.

Figures~\ref{fig:ad_mdot} and~\ref{fig:bc_mdot} show how  
the Eddington ratio affects the SED and the bolometric correction. As the Eddington ratio increases, the minimum of the bolometric correction function, corresponding to the peak of the accretion disk emission, moves towards higher frequencies.
At frequencies below the minimum, the sources with lower Eddington ratio require a smaller correction than their higher Eddington ratio counterparts because the peak of their accretion disk occurs at lower frequencies. In contrast, at frequencies above than the minimum, the opposite happens.  In the X-ray bands, the accretion disk templates converge for the same BH mass and spin
(Figure~\ref{fig:ad_mdot}).  This is inherent in the QSOSED model and is due to a combination of factors: an increase in Eddington ratio resulting in an increase in luminosity \citep{Ho2008} and a steepening of $\alpha_{ox}$ and $\Gamma$(2--10 keV) (For more details see \citealt{Kubota2018}).  
Because $\nu L_{\nu}$ in the X-ray band varies little with Eddington ratio (Figure \ref{fig:ad_mdot}), 
the strong dependence of the accretion disk luminosity on Eddington ratio results in  
a larger range of bolometric correction 
in the X-ray (Figure~\ref{fig:bc_mdot}). 
As BH mass increases, the bolometric correction at frequencies below the minimum decreases due to the peak of the accretion disk emission moving towards lower frequencies. For the same reason,  at X-ray frequencies, a higher bolometric correction is needed for a higher BH mass.

Figures~\ref{fig:ad_mass}--\ref{fig:bc_mdot} also reveal how varying BH spin affects the accretion disk SED and the bolometric correction.
Radio-loud AGN are likely to have BHs with high spin in order to power the radio jets. However, radio-quiet AGN may also have high spin values \citep{lb2006}. In practice, small changes in spin do not significantly change the accretion disk SED \citep[see Figure 1 in][]{Azadi2020}, thus we include spins 0 and 1 for illustrative purposes. The spin values considered here cover the widest possible range of AGN from radio-silent to radio-loud. For a given BH mass and Eddington ratio, a rapidly rotating black hole has a hotter accretion disk than its low-spin counterpart. This results in a bolometric correction dip at higher frequencies as spin increases. At frequencies lower than the minimum, a high-spin BH requires a larger bolometric correction than a low-spin BH, while the opposite happens above the minimum. This is because an increase in spin moves the peak of the accretion disk emission towards higher energies. 

In summary, Figures~\ref{fig:ad_mass} to \ref{fig:bc_mdot} show how BH mass, Eddington ratio and spin impact the accretion disk SED and bolometric correction. An observer with photometry ---corrected for extinction--- in the range from 1\,\um to 10\,keV can apply the SEDs from Figures~\ref{fig:ad_mass} and~\ref{fig:ad_mdot} to replicate the accretion disk emission of radio-silent or radio-loud quasar at $z\sim1.5$. Then the bolometric power of any quasar at this redshift can be estimated using the bolometric correction factors from Figures~\ref{fig:bc_mass} or~\ref{fig:bc_mdot}.

\begin{figure*}[th!]
\begin{center}
    \includegraphics[height=\textwidth,angle =90]{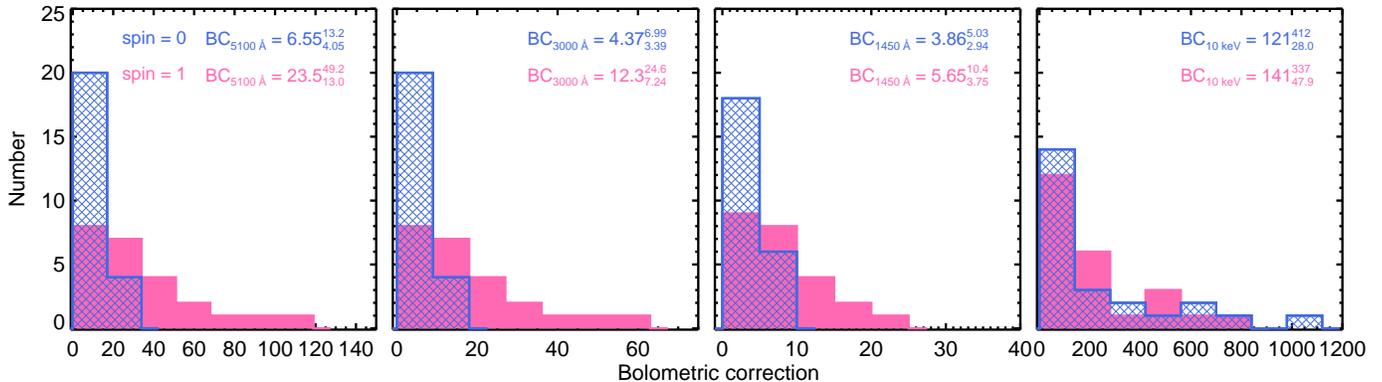}
 \caption{The distributions of  bolometric corrections at 5100\,\AA, 3000\,\AA, 1450\,\AA\ and 10\,keV. Non-spinning (i.e., spin = 0) models are shown in blue, and maximally-spinning (i.e., spin = 1) models are shown in pink. Distributions are based on 
the median bolometric correction curves in Figures \ref{fig:bc_mass} and \ref{fig:bc_mdot}, thus depend on BH mass, Eddington ratio and spin. The median values for each distribution and the corresponding 25th--75th percentile envelopes are given in each panel.}    
\label{fig:bc_hist}
\end{center}
\end{figure*}

\begin{table*}[th!]
    \caption{Bolometric correction factors compared}
    \begin{tabular}{cccccccc}
    \hline \hline
        Wavelength/Energy &\multicolumn{2}{c}{This paper}  & \citeauthor{Elvis1994}   &\citeauthor{Richards2006} &\citeauthor{Nemmen2010}  &\citeauthor{R12} \\
        (rest-frame)& spin = 0  & spin = 1 &(1994)&(2006)&(2010)&(2012) \\
         \hline 
        \vspace{0.01pt}\\
5100~\AA
&$6.55^{13.2}_{4.05}$ & $23.5^{49.2}_{13.0}$
&12.7 $(7.68\tablenotemark{a})$&$10.3\pm2.1$&7.6&8.1\vspace{0.15cm}\\
3000~\AA 
&$4.37^{6.99}_{3.39}$ & $12.3^{24.6}_{7.24}$
&$(3.82\tablenotemark{a})$&
$5.62\pm1.14$&5.9
&5.2\vspace{0.15cm}\\
1450~\AA&
$3.86^{5.03}_{2.94}$ & $5.65^{10.4}_{3.75}$
&$ (3.15\tablenotemark{a})$
& $(2.33\tablenotemark{b})$&3.0&4.2\vspace{0.15cm}\\
10~keV&
$121^{412}_{28.0}$ & $141^{337}_{47.9}$
&\nodata&\nodata&\nodata&38.0\0\vspace{0.15cm}\\
\hline
    \end{tabular}
    \label{tab:wave_bc}
\tablenotetext{a} {values from integrating the mean SED of the radio-quiet quasars of  \citet{Elvis1994} from 1\,\micron\ to 8\,keV as reported by \citet{R12}.  }
\tablenotetext{b} {\citet{Richards2006} did not report the correction at 1450\,\AA. However, \cite{R12} reported the value for the mean SED of all 
\citeauthor{Richards2006} quasars  integrating from 1\,\um to 8\,keV.}
\end{table*}

\subsection{Comparison With the Literature at Visible Wavelengths} \label{sec:lit}

Here we compare our estimated bolometric correction factors at different wavelengths with four commonly used values from the literature. Figure~\ref{fig:bc_hist} shows the distributions of the bolometric corrections calculated at 5100\,\AA, 3000\,\AA, 1450\,\AA, and 10\,keV for non-rotating BH (i.e., spin= 0) and maximally-rotating BH (i.e., spin= 1) quasars. To determine the bolometric correction at each wavelength, we considered the median curves from Figures~\ref{fig:bc_mass} and \ref{fig:bc_mdot}.  Our bolometric correction estimates therefore reflect variations in BH mass, Eddington ratio, and spin.

Given that 1450\,\AA\ is closest to the wavelength at which the accretion-disk SED peaks, smaller bolometric corrections are required to infer the quasars total luminosities from their 1450\,\AA\ emission than from other photometry. At any of these, wavelengths maximally-rotating BH require a higher correction than their non-rotating counterparts. This is because the innermost stable orbit of the accretion disk for spinning BHs lies closer to the BH, resulting in a more luminous accretion disk at UV wavelengths (see also Figure~\ref{fig:ad_mass}). 
The distributions for spinning versus non-spinning
differ at the $4\sigma$, $4\sigma$, $>2\sigma$ and $>1\sigma$ levels at 5100\,\AA, 3000\,\AA, 1450\,\AA, and 10\,keV, respectively, according to a Kolmogorov--Smirnov test. 

Table \ref{tab:wave_bc} compares the 
bolometric corrections estimated here 
with those from the literature. Broadly
speaking, the estimates agree well. \cite{Richards2006} estimated the bolometric correction for a sample of SDSS quasars for which multi-wavelength radio-to-X-ray data are available. The sample is dominated by radio-quiet quasars and includes both blue and red (intrinsically reddened) Type~1 quasars.  
Our median values for radio-quiet quasars are consistent within the uncertainties with \cite{Richards2006}.
Slight differences between our results and those of
\citeauthor{Richards2006} may have arisen because \citeauthor{Richards2006} integrated over 100\,\um--10\,keV while our integration spans the range 1\,\um--10\,keV and does not include reprocessed emission in the IR\null. Their sample also includes both radio-loud and -quiet quasars and does not distinguish them. 

\cite{Elvis1994} analyzed a $z<1$, heterogeneous (half radio and half optically selected) sample of bright quasars observed by the {Einstein Observatory} with sufficient counts to constrain the soft-X-ray slope.
The sample was also chosen to be detected by {IUE} and have sufficient multi-wavelength (radio to X-ray) photometry available. Therefore, their sample is biased toward blue (unobscured) quasars with strong soft-X-ray emission. 
As shown in Table \ref{tab:wave_bc} our bolometric correction estimations broadly agree with the
\cite{Elvis1994} values. However, \cite{Elvis1994} considered a wider range (IR--X-ray) of integration than ours; additionally the sample selection and radio heterogeneity result in different bolometric correction than ours.

\cite{R12} studied a sample of 63 bright (radio-loud and -quiet) quasars at $z<1.4$. Unlike our results, \cite{R12} found similar (within 95\% confidence intervals) bolometric corrections for radio-quiet and -loud populations at visible--UV wavelengths, and found a significant difference only in the X-ray regime.  While \cite{R12} integrated over a wavelength range similar to ours (1\,\um--8\,keV), they used observed SEDs and covered the gap in observations with extrapolations, whereas our models self-consistently cover the entire range. 

Unlike \cite{Elvis1994}, \cite{Richards2006}, and \cite{R12}, who determined empirical bolometric corrections from observed SEDs, \cite{Nemmen2010} used the \cite{Hubeny2001} accretion disk model to derive theoretical bolometric corrections.   
However, the \cite{Nemmen2010} model extends only from 3\,\um to 414\,eV  
and  does not account for hard-X-ray emission from the corona.


\begin{figure*}[t!]
    \includegraphics[width=0.40 \textwidth,angle =90]{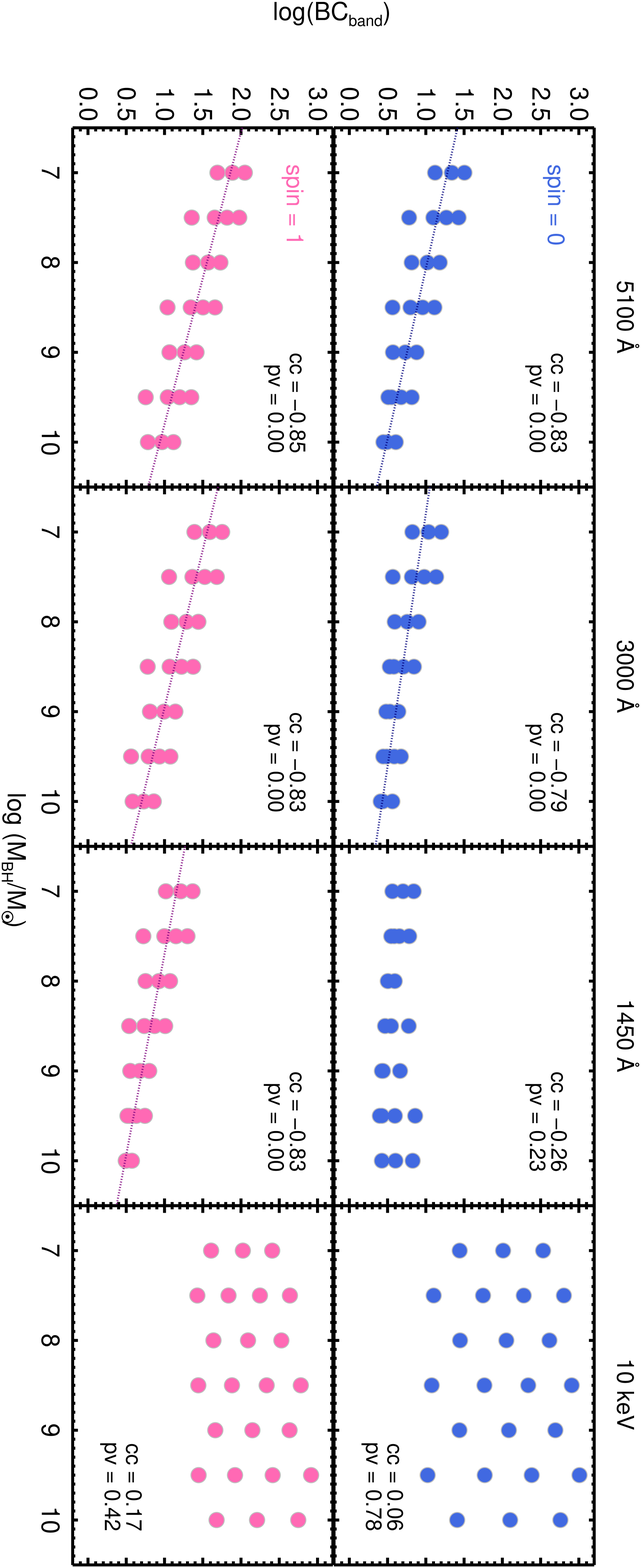}
     \caption{The dependence of the bolometric correction (BC) on BH mass. Panels from left to right show wavelength or energy as labeled above the panels. Top row shows values for non-rotating (spin=0), and bottom row shows values for maximally-rotating (spin=1). For each mass, the spread in bolometric correction is due to the range of -1.5 to 0.0 for $\log(\lambda_{Edd})$. The correlation coefficient (cc) and the $p$-value(pv) of the correlations are shown in each panel, and dotted lines show the best fit  when a significant correlation is present. Table~\ref{tab:eq_mass} gives coefficients for the line equations.}
\label{fig:wave_mass}
\vspace{0.5cm}
\includegraphics[width=0.40\textwidth,angle =90]{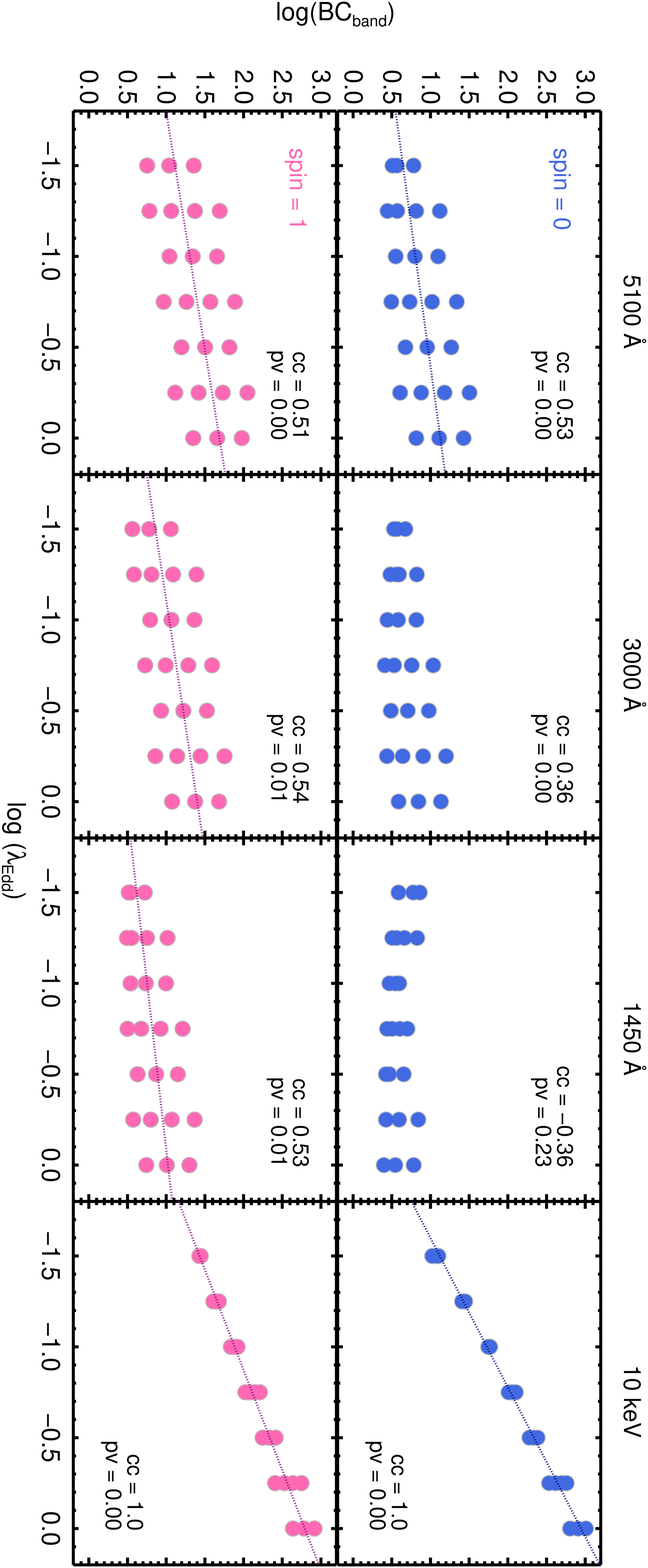}
    \caption{{\sl From left to right:} The dependence of the bolometric correction (BC) at 5100\AA, 3000\AA, 1450\AA\ and 10~keV, on Eddington ratio.  Panels from left to right show wavelength or energy as labeled above the panels. Top row shows values for non-rotating (spin=0), and bottom row shows values for maximally-rotating (spin=1). For each $\log (\lambda_{Edd})$, the spread in bolometric correction is due to the range of $10^{7}$ to $10^{10}$ of BH mass. Correlation coefficient (cc) and $p$-value (pv) are shown in each panel, and  dotted lines show the best fit when a significant correlation is present. Table~\ref{tab:eq_mdot} gives coefficients for the line equations.}    
\label{fig:wave_mdot}
\end{figure*}

\begin{table*}
\centering
    \caption{The best fitted line equations in Figures \ref{fig:wave_mass}}
    \begin{tabular}{cccc}
    \hline \hline
        Wavelength/Energy & spin = 0 &  spin = 1\\
         \hline 
5100~\AA
&$\log({\rm BC}) = (-0.26\pm0.03) \log (\frac { M_{\rm BH}}{ 10^9M_{\odot}} )+  (0.79\pm0.03)$
&$\log({\rm BC}) = (-0.31\pm0.04) \log(\frac { M_{\rm BH}}{ 10^9M_{\odot}} ) +  (1.22\pm0.04)\vspace{0.15cm}$\\
3000~\AA 
&$\log({\rm BC}) = (-0.18\pm0.03)\log(\frac { M_{\rm BH}}{ 10^9M_{\odot}} )+  (0.59\pm0.03)$
&$\log({\rm BC}) = (-0.28\pm0.04)\log(\frac { M_{\rm BH}}{ 10^9M_{\odot}} ) +  (1.03\pm0.04)\vspace{0.15cm}$\\
1450~\AA
&
\nodata&$\log({\rm BC}) = (-0.22\pm0.03) \log(\frac { M_{\rm BH}}{ 10^9M_{\odot}} ) +  (0.74\pm0.03)\vspace{0.15cm}$\\
10~keV
&
\nodata&\nodata \vspace{0.15cm}\\
\hline
    \end{tabular}
    \label{tab:eq_mass}
\end{table*} 

\begin{table*}
\centering
\small
\caption{The best fitted line equations in Figure \ref{fig:wave_mdot}}
\begin{tabular}{cccc}
\hline \hline
Wavelength/Energy & spin = 0 &  spin = 1\\
 \hline 
5100~\AA
&$\log({\rm BC}) = (0.32\pm0.12) \log(\rm \lambda_{Edd}) +  (1.13\pm0.12)$
&$\log({\rm BC}) = (0.38\pm0.13) \log(\rm \lambda_{Edd}) +  (1.69\pm0.13)\vspace{0.15cm}$\\
3000~\AA 
&\nodata
&$\log({\rm BC}) = (0.36\pm0.12) \log(\rm \lambda_{Edd}) +  (1.40\pm0.12)\vspace{0.15cm}$\\
1450~\AA&
\nodata&$\log({\rm BC}) = (0.27\pm0.01) \log(\rm \lambda_{Edd}) +  (1.02\pm0.01)\vspace{0.15cm}$\\
10~keV&
$\log({\rm BC}) = 
(1.22\pm0.03)
\log(\rm \lambda_{Edd}) +  
(2.94\pm0.03)$
&$\log({\rm BC}) = 
(0.91\pm0.03) \log(\rm 
\lambda_{Edd}) +  
(2.79\pm0.03)$ \\[0.15cm]
\hline

    \end{tabular}
    \label{tab:eq_mdot}
\end{table*}

\subsection{Comparison with the Literature at X-ray Wavelengths} \label{sec:lit2}

We estimate 10\,keV median correction factors of 121 and 141 for non-spinning and maximally-spinning BHs, respectively.  \cite{R12} estimated bolometric corrections of $23$ and $89$, respectively, for the radio-loud and -quiet samples (and 38 for the full sample) based on the assumption of isotropy. \cite{R12} adjusted the luminosity for different viewing angle assuming $L_{bol}\approx0.75\times L_{\rm isotropy}$, where the factor of 0.75 accounts for anisotropic accretion disk emission for the assumed viewing angle.
While \citeauthor{R12} determined a higher bolometric correction for radio-quiet sources than their radio-loud counterparts, our SED templates in Figures~\ref{fig:ad_mass} and~\ref{fig:ad_mdot} indicate a higher spin results in larger corrections. Although the \cite{R12} estimate is consistent with ours within the uncertainties, our median values for both non-spinning and maximally-spinning populations are higher. \cite{R12} covered the gap between FUV and X-ray observations with a linear interpolation using either a power law or \cite{FM1987} and \cite{K1997} models. These models result in a higher luminosity than our model at these wavelengths and consequently a smaller bolometric correction than ours.

\cite{Hopkins2007} presented the bolometric luminosity function from the MIR to the hard X-ray regime for quasars at $z<6$ and their luminosity-dependent bolometric correction.  For quasars with 2\,keV luminosities of $10^{44}$\,erg\,s$^{-1}$, they found a bolometric correction of $\sim$20, while we estimate a bolometric correction of  $\sim$50--200 depending on the BH mass, Eddington ratio, and spin.  
\cite{Hopkins2007} primarily approximated the SED based on the visible-to-X-ray spectral index ($\alpha_{\rm ox}$) 
and  $L_{\nu} (2500\,{\rm \AA})$ 
relation and did not take the intrinsic differences in  accretion-disk SEDs resulting from  BH mass, Eddington ratio, and spin into account (See also \citealt{V2007}).

Based on an analysis of $z<0.7$ AGN with a combination of Far Ultraviolet Spectroscopic Explorer (FUSE) UV data and X-ray photometry from the Advanced Satellite for Cosmology and Astrophysics (ASCA),  {XMM–Newton}, and {Chandra},
\cite{V2007} found no strong evidence of a luminosity dependence for the bolometric correction but did find an Eddington-ratio dependence. Specifically, \citeauthor{V2007} found that for $\log(\lambda_{\rm Edd})<-1.0$, the bolometric correction at X-ray is typically 15--25, while above that $\lambda_{\rm Edd}$, it is typically 40--70. Our findings 
are generally consistent with \cite{V2007}, and as  discussed below in Section~\ref{sec:wavelengths}, we find a strong dependence of the bolometric correction at X-ray to the Eddington ratio.

\subsection{Dependence of the 5100\,\AA, 3000\,\AA, 1450\,\AA\ and 10\,keV 
Bolometric Corrections on AGN Properties} \label{sec:wavelengths}

Figures~\ref{fig:wave_mass} and \ref{fig:wave_mdot} illustrate the dependence of the bolometric corrections at 5100\,\AA, 3000\,\AA, 1450\,\AA\ and 10\,keV on the BH mass, Eddington ratio, and spin. The data points in these plots are from the median curves in Figures~\ref{fig:bc_mass} and \ref{fig:bc_mdot}.  
To quantify the relation between the bolometric correction and BH mass or Eddington ratio, we calculated the correlation coefficient and the corresponding significance using IDL's {\tt r-correlate} routine, which computes the Spearman rank correlation coefficient (cc) and the significance of its deviation from zero (pv). For cases with 
significance ${>}3\sigma$, the relation is shown in the figure, and the fitted line parameters are given in Tables~\ref{tab:eq_mass} and~\ref{tab:eq_mdot}.

In Figure~\ref{fig:wave_mass}, in the 5100\,\AA\ and 3000\,\AA\ panels and 1450\,\AA\ for maximally-spinning sources, a low-mass source requires a larger bolometric correction than a high-mass one. Also, the data points at lower masses cover a wider range than at higher masses. There is a statistically significant correlation between the bolometric correction and the BH mass for both spins (See Table~\ref{tab:eq_mass}). However, in the non-spinning population, the relation disappears at 1450\,\AA. As BH mass increases, the peak of accretion-disk emission moves towards longer wavelengths, resulting in a statistically significant relation between the two parameters at 5100 and 3000\,\AA, while at 1450\,\AA\, close to the peak of the accretion disk SED, the mass dependence disappears in non-spinning sources. However, this is not the case in the maximally-spinning population, given that the higher spin of these sources pushes the peak to higher frequencies. Figure~\ref{fig:wave_mass} shows that the 10\,keV bolometric correction is independent of BH mass for both spin values. This is expected because 10\,keV emission arises primarily from the corona, not the accretion disk.

Figure~\ref{fig:wave_mdot} shows the dependence of the bolometric correction on the Eddington ratio. At 5100 and 3000\,\AA\, the bolometric correction increases with Eddington ratio (for both spin values). An increase in the Eddington ratio moves the peak to higher frequencies, resulting in a larger correction at higher Eddington ratios. Again, at 1450\,\AA, there is no strong dependence on the Eddington ratio for non-spinning sources because 1450\,\AA\ is closer to the peak of the accretion-disk emission, while for maximally spinning sources, there is a weak (but statistically significant) relation between the two. Figures~\ref{fig:wave_mass} and~\ref{fig:wave_mdot} show the dependence of the  bolometric corrections on Eddington ratio is weaker compared to dependence on mass at visible bands. However, at 10\,keV, there is a strong dependence of bolometric correction on  Eddington ratio, as the coronal emission varies strongly with the Eddington ratio (Table~\ref{tab:eq_mdot}).

In summary, our analysis reveals a strong dependence of the bolometric correction on BH mass in the visible bands and a  strong dependence on the Eddington ratio at hard X-ray energies. The dependence varies slightly from non-spinning to maximally-spinning populations. However, the dependence becomes weaker closer to the peak of the accretion disk SED.

%% file: summary.tex
\section{Summary} \label{sec:summary}

To better understand how AGN impact their host galaxies and nearby environments (i.e., galaxy clusters), it is critical to determine their total radiative power. In this study we have presented the accretion disk SEDs and bolometric correction factors for radio-quiet to radio-loud quasars at $z\sim1.5$. 

\begin{itemize}

\item We present accretion disk SEDs of non-rotating (spin = 0) to maximally-rotating (spin =1) BHs in quasars using \citep{Kubota2018} QSOSED model. Our accretion disk templates cover ranges of mass:  $10^7$ to $10^{10}$~\Msun,  Eddington ratio:  $0.03<\log(\lambda_{Edd})<1$, and spin: 0--1. The viewing angle in our model is assumed to be 30$^{\circ}$ which is appropriate for quasars. These templates can replicate the accretion disk emission in any sample of radio-quiet to radio-loud Type~1 AGN (Figures \ref{fig:ad_mass}, \ref{fig:ad_mdot} and Tables \ref{tab:fig5},\ref{tab:fig7}).

\item To estimate the bolometric luminosity of AGN, we have integrated  accretion disk SEDs of \citet{Kubota2018} from 1\,\micron\ to 10\,keV. There is no need to include the MIR emission because reprocessed  photons are accounted for in the visible--UV portion of the SED. No further correction for the viewing angle is needed. Additionally, no gap repair is required because our model self-consistently covers any gap in observations (Equation~\ref{eq:bc}).

\item The accretion disk SED and consequently the bolometric corrections inferred from it strongly depend on  BH mass, spin, and Eddington ratio (Figures~\ref{fig:bc_mass} and~\ref{fig:bc_mdot} and Tables~\ref{tab:fig6} and~\ref{tab:fig8}). 

\item As expected, the smallest bolometric correction is for frequencies close to the peak of the accretion-disk emission. The peak-emission frequency decreases with BH mass (if Eddington ratio and spin are held constant), and therefore the frequency where the bolometric correction is minimized decreases as BH mass increases (Figures~\ref{fig:bc_mass}, \ref{fig:bc_mdot}). 

\item Increasing the Eddington ratio (while holding BH mass and spin constant) moves the peak of the accretion disk SED to higher frequencies.  Thus, at higher frequencies a lower correction is required to infer the bolometric luminosity.  At lower frequencies the opposite happens and higher bolometric corrections are required for higher Eddington ratios (Figures \ref{fig:bc_mass}--\ref{fig:bc_mdot}). 

\item Maximally-rotating (spin = 1) quasars require a higher bolometric correction than non-rotating (spin = 0) quasars at all wavelengths. A maximally-rotating quasar of a given BH mass and Eddington ratio has a more luminous and hotter accretion disk than its non-spinning counterparts. (Figures \ref{fig:bc_mass}--\ref{fig:bc_mdot}).

\item For non-spinning BHs, the median bolometric corrections (and 25th--75th percentiles) are consistent with published values, although previous studies do not present corrections as a function of spin (Figure~\ref{fig:bc_hist}).

\item The bolometric correction  in the visible bands (5000\,\AA\ and 3000\,\AA) depends strongly on BH mass while at hard X-ray energies it strongly  
depends on Eddington ratio.  However, both relationships are weaker closer to the peak (in UV) of the accretion disk SED (Figures \ref{fig:wave_mass} and \ref{fig:wave_mdot}).

\item The bolometric corrections of \z 3CRR quasars show no evidence for a  dependence on redshift or \nh, but the  3CRR quasar sample used here offers only a limited range of these properties (Figure \ref{fig:agn_prop}).

\end{itemize}

The SED templates presented here are applicable to any Type 1 quasar at $z\sim1.5$.  An observer with photometry in the range from 1\,\um to 10\,keV can apply these SEDs and their corresponding bolometric correction to estimate the  intrinsic radiative power of their sources provided the observations can be corrected for extinction or such correction is negligible.

\section*{Acknowledgements} 

Support for this work was provided by NASA grants: \#80NSSC18K1609,  \#80NSSC19K1311,
 \#80NSSC20K0043  (MAz), and \#80NSSC21K0058 (JK), and by NASA Contract NAS8-03060   \textit{Chandra} X-ray Center (CXC), which is operated by the Smithsonian Astrophysical Observatory   
(BJW, JK). 

The scientific results in this article are based to a significant degree on observations made by the \textit{Chandra}~X-ray Observatory (CXO).
This research has made use of data obtained from the \textit{Chandra} Data
Archive. This research is based on observations made by {\it Herschel}, which
is an ESA space observatory with science instruments provided by European-led Principal Investigator consortia and with important
participation from NASA. This work is based in part on observations made with the Spitzer Space
Telescope, which was operated by the Jet Propulsion Laboratory, California Institute of Technology under a contract with NASA.

We acknowledge the use of Ned Wright's calculator
\citep{2006PASP..118.1711W} and NASA/IPAC Extragalactic Database (NED), operated by the Jet Propulsion Laboratory, California Institute
of Technology, under contract with the National Aeronautics and Space
Administration.

The authors would like to thank Chris Done, Malgosia Sobolewska, Mark Birkinshaw and Diana Worrall for helpful comments that improved the quality of the paper.

%% file: Appendix.tex
\FloatBarrier
\appendix
Tables \ref{tab:fig5} to \ref{tab:fig8} below tabulated the SED plots of Figure \ref{fig:ad_mass} and \ref{fig:ad_mdot} as well as the bolometric correction of Figure \ref{fig:bc_mass} and \ref{fig:bc_mdot}.
\setlength{\tabcolsep}{0.5em}
\null\vspace{-4.5in}
\begin{sidewaystable}
\renewcommand{\arraystretch}{1.2}
\tablenum{A1}
\caption{Accretion disk SED 
as a function of black hole mass (Figure \ref{fig:ad_mass})}

\tiny
\hspace{-0.5in}
\begin{tabular}{c@{\hspace{2em}}ccc@{\hspace{3em}}ccc@{\hspace{3em}}ccc@{\hspace{3em}}ccc}
    \hline\hline
    \textbf{spin = 0}
    &\multicolumn{3}{c}{$\log (\lambda_{Edd}) = -1.5$}
    &\multicolumn{3}{c}{$\log (\lambda_{Edd}) = -1.0$}
    &\multicolumn{3}{c}{$\log (\lambda_{Edd}) = -0.5$}
    &\multicolumn{3}{c}{$\log (\lambda_{Edd}) = 0.0$}\\
          \hline 
          $M_{\rm BH}=$
          &$10^7$--$10^8$
          &$10^8$--$10^9$
          &$10^9$--$10^{10}$
          &$10^7$--$10^8$
          &$10^8$--$10^9$
          &$10^9$--$10^{10}$
          &$10^7$--$10^8$
          &$10^8$--$10^9$
          &$10^9$--$10^{10}$
          &$10^7$--$10^8$
          &$10^8$--$10^9$
          &$10^9$--$10^{10}$\\
          \hline 
          $\log{\nu}$\\
14.00&$42.7^{43.4}_{42.1}$&$44.1^{44.6}_{43.4}$&$45.3^{45.8}_{44.7}$&$43.1^{43.7}_{42.4}$&$44.5^{45.1}_{43.8}$&$45.8^{46.4}_{
45.2}$&$43.5^{44.1}_{42.8}$&$44.9^{45.5}_{44.2}$&$46.2^{46.8}_{45.5}$&$43.9^{44.5}_{43.2}$&$45.2^{45.8}_{44.5}$&$46.6^{47.2}_{
45.9}$\\
14.25&$42.7^{43.4}_{42.1}$&$44.1^{44.6}_{43.4}$&$45.3^{45.8}_{44.7}$&$43.1^{43.7}_{42.4}$&$44.5^{45.1}_{43.8}$&$45.8^{46.4}_{
45.2}$&$43.5^{44.1}_{42.8}$&$44.9^{45.5}_{44.2}$&$46.2^{46.8}_{45.5}$&$43.9^{44.5}_{43.2}$&$45.2^{45.8}_{44.5}$&$46.6^{47.2}_{
45.9}$\\
14.50&$42.9^{43.5}_{42.3}$&$44.3^{44.8}_{43.6}$&$45.4^{45.9}_{44.9}$&$43.3^{43.9}_{42.6}$&$44.7^{45.3}_{44.0}$&$46.0^{46.6}_{
45.3}$&$43.7^{44.3}_{43.0}$&$45.0^{45.6}_{44.4}$&$46.4^{46.9}_{45.7}$&$44.0^{44.7}_{43.4}$&$45.4^{46.0}_{44.7}$&$46.7^{47.3}_{
46.1}$\\
14.75&$43.4^{43.9}_{42.8}$&$44.6^{45.1}_{44.0}$&$45.6^{46.0}_{45.1}$&$43.8^{44.3}_{43.1}$&$45.0^{45.6}_{44.4}$&$46.3^{46.8}_{
45.7}$&$44.1^{44.7}_{43.4}$&$45.4^{46.0}_{44.8}$&$46.7^{47.2}_{46.0}$&$44.5^{45.0}_{43.8}$&$45.8^{46.3}_{45.1}$&$47.1^{47.6}_{
46.4}$\\
15.00&$43.6^{44.1}_{43.0}$&$44.6^{45.0}_{44.1}$&$45.4^{45.8}_{45.0}$&$44.1^{44.6}_{43.4}$&$45.3^{45.8}_{44.7}$&$46.4^{46.8}_{
45.8}$&$44.4^{45.0}_{43.8}$&$45.7^{46.2}_{45.0}$&$46.9^{47.4}_{46.3}$&$44.8^{45.3}_{44.1}$&$46.1^{46.6}_{45.4}$&$47.3^{47.8}_{
46.7}$\\
15.25&$43.6^{44.0}_{43.2}$&$44.4^{44.8}_{44.0}$&$45.3^{45.7}_{44.8}$&$44.3^{44.8}_{43.7}$&$45.4^{45.8}_{44.9}$&$46.3^{46.6}_{
45.9}$&$44.7^{45.2}_{44.0}$&$45.9^{46.4}_{45.3}$&$46.9^{47.3}_{46.4}$&$45.1^{45.6}_{44.4}$&$46.3^{46.8}_{45.7}$&$47.5^{47.9}_{
46.9}$\\
15.50&$43.4^{43.8}_{43.1}$&$44.3^{44.7}_{43.8}$&$45.2^{45.6}_{44.7}$&$44.4^{44.8}_{43.9}$&$45.3^{45.6}_{44.9}$&$46.0^{46.4}_{
45.7}$&$44.9^{45.4}_{44.3}$&$45.9^{46.4}_{45.4}$&$46.8^{47.1}_{46.4}$&$45.3^{45.8}_{44.7}$&$46.5^{46.9}_{45.9}$&$47.5^{47.8}_{
47.0}$\\
15.75&$43.3^{43.7}_{42.8}$&$44.2^{44.6}_{43.7}$&$45.1^{45.5}_{44.6}$&$44.3^{44.6}_{43.9}$&$45.0^{45.4}_{44.7}$&$45.9^{46.3}_{
45.4}$&$44.9^{45.4}_{44.4}$&$45.8^{46.1}_{45.4}$&$46.5^{46.8}_{46.1}$&$45.5^{45.9}_{44.9}$&$46.5^{46.8}_{46.0}$&$47.2^{47.4}_{
46.9}$\\
16.00&$43.2^{43.6}_{42.8}$&$44.1^{44.5}_{43.6}$&$45.0^{45.4}_{44.5}$&$44.1^{44.4}_{43.7}$&$44.9^{45.3}_{44.5}$&$45.8^{46.2}_{
45.3}$&$44.8^{45.1}_{44.4}$&$45.5^{45.8}_{45.1}$&$46.3^{46.7}_{45.9}$&$45.5^{45.8}_{45.0}$&$46.2^{46.5}_{45.9}$&$46.8^{47.2}_{
46.5}$\\
16.25&$43.1^{43.5}_{42.7}$&$44.0^{44.4}_{43.6}$&$44.9^{45.3}_{44.5}$&$43.9^{44.3}_{43.5}$&$44.8^{45.2}_{44.4}$&$45.7^{46.1}_{
45.2}$&$44.5^{44.9}_{44.2}$&$45.3^{45.7}_{44.9}$&$46.2^{46.6}_{45.8}$&$45.2^{45.5}_{44.9}$&$45.9^{46.2}_{45.5}$&$46.7^{47.1}_{
46.3}$\\
16.50&$43.0^{43.4}_{42.6}$&$43.9^{44.4}_{43.5}$&$44.9^{45.3}_{44.4}$&$43.8^{44.2}_{43.4}$&$44.7^{45.1}_{44.3}$&$45.6^{46.0}_{
45.1}$&$44.3^{44.7}_{43.9}$&$45.2^{45.6}_{44.8}$&$46.1^{46.5}_{45.6}$&$44.9^{45.2}_{44.5}$&$45.7^{46.1}_{45.3}$&$46.6^{46.9}_{
46.1}$\\
16.75&$43.0^{43.4}_{42.5}$&$43.9^{44.3}_{43.4}$&$44.8^{45.3}_{44.4}$&$43.7^{44.1}_{43.3}$&$44.6^{45.0}_{44.1}$&$45.5^{45.9}_{
45.0}$&$44.2^{44.6}_{43.8}$&$45.1^{45.5}_{44.6}$&$45.9^{46.3}_{45.5}$&$44.7^{45.1}_{44.3}$&$45.6^{45.9}_{45.1}$&$46.4^{46.8}_{
46.0}$\\
17.00&$42.9^{43.3}_{42.4}$&$43.9^{44.3}_{43.4}$&$44.8^{45.2}_{44.3}$&$43.6^{44.0}_{43.1}$&$44.5^{44.9}_{44.0}$&$45.4^{45.8}_{
44.9}$&$44.0^{44.4}_{43.6}$&$44.9^{45.3}_{44.5}$&$45.8^{46.2}_{45.3}$&$44.5^{44.9}_{44.1}$&$45.4^{45.8}_{45.0}$&$46.2^{46.6}_{
45.8}$\\
17.25&$42.9^{43.3}_{42.4}$&$43.8^{44.3}_{43.4}$&$44.8^{45.3}_{44.3}$&$43.4^{43.8}_{42.9}$&$44.3^{44.7}_{43.9}$&$45.2^{45.7}_{
44.8}$&$43.8^{44.2}_{43.4}$&$44.7^{45.1}_{44.2}$&$45.6^{46.0}_{45.1}$&$44.3^{44.7}_{43.9}$&$45.1^{45.5}_{44.7}$&$46.0^{46.4}_{
45.5}$\\
17.50&$42.9^{43.3}_{42.4}$&$43.8^{44.3}_{43.3}$&$44.8^{45.3}_{44.3}$&$43.2^{43.6}_{42.7}$&$44.1^{44.6}_{43.7}$&$45.1^{45.5}_{
44.6}$&$43.5^{43.9}_{43.0}$&$44.4^{44.8}_{43.9}$&$45.3^{45.7}_{44.8}$&$43.8^{44.2}_{43.4}$&$44.7^{45.1}_{44.2}$&$45.5^{45.9}_{
45.1}$\\
17.75&$42.9^{43.3}_{42.4}$&$43.9^{44.3}_{43.4}$&$44.9^{45.3}_{44.4}$&$43.1^{43.6}_{42.6}$&$44.1^{44.5}_{43.6}$&$45.1^{45.5}_{
44.6}$&$43.2^{43.7}_{42.8}$&$44.2^{44.6}_{43.7}$&$45.1^{45.5}_{44.6}$&$43.4^{43.8}_{42.9}$&$44.3^{44.7}_{43.8}$&$45.2^{45.6}_{
44.7}$\\
18.00&$43.0^{43.4}_{42.5}$&$44.0^{44.4}_{43.5}$&$45.0^{45.4}_{44.5}$&$43.1^{43.5}_{42.6}$&$44.1^{44.5}_{43.6}$&$45.1^{45.5}_{
44.6}$&$43.2^{43.6}_{42.7}$&$44.1^{44.5}_{43.6}$&$45.0^{45.5}_{44.6}$&$43.3^{43.7}_{42.8}$&$44.1^{44.5}_{43.7}$&$45.0^{45.4}_{
44.6}$\\
18.25&$43.0^{43.5}_{42.5}$&$44.0^{44.5}_{43.5}$&$45.0^{45.5}_{44.5}$&$43.1^{43.5}_{42.6}$&$44.1^{44.5}_{43.6}$&$45.1^{45.5}_{
44.6}$&$43.1^{43.5}_{42.7}$&$44.1^{44.5}_{43.6}$&$45.0^{45.4}_{44.5}$&$43.1^{43.5}_{42.7}$&$44.0^{44.4}_{43.6}$&$44.9^{45.3}_{
44.5}$\\
18.50&$43.1^{43.5}_{42.6}$&$44.1^{44.5}_{43.6}$&$45.1^{45.5}_{44.6}$&$43.1^{43.5}_{42.6}$&$44.1^{44.5}_{43.6}$&$45.1^{45.5}_{
44.6}$&$43.1^{43.5}_{42.6}$&$44.0^{44.4}_{43.6}$&$45.0^{45.4}_{44.5}$&$43.1^{43.5}_{42.6}$&$44.0^{44.4}_{43.5}$&$44.9^{45.3}_{
44.4}$\\[0.5ex]
         \hline
          \textbf{spin = 1}
    \\[0.5ex]
14.00&$42.6^{43.2}_{41.9}$&$44.0^{44.6}_{43.3}$&$45.3^{45.9}_{44.6}$&$42.8^{43.4}_{42.1}$&$44.2^{44.8}_{43.5}$&$45.5^{46.1}_{
44.8}$&$43.1^{43.7}_{42.4}$&$44.5^{45.1}_{43.8}$&$45.8^{46.4}_{45.2}$&$43.5^{44.1}_{42.8}$&$44.8^{45.4}_{44.2}$&$46.2^{46.8}_{
45.5}$\\
14.25&$42.6^{43.2}_{41.9}$&$44.0^{44.6}_{43.3}$&$45.3^{45.9}_{44.6}$&$42.8^{43.4}_{42.1}$&$44.2^{44.8}_{43.5}$&$45.5^{46.1}_{
44.8}$&$43.1^{43.7}_{42.4}$&$44.5^{45.1}_{43.8}$&$45.8^{46.4}_{45.2}$&$43.5^{44.1}_{42.8}$&$44.8^{45.4}_{44.2}$&$46.2^{46.8}_{
45.5}$\\
14.50&$42.8^{43.4}_{42.1}$&$44.1^{44.7}_{43.5}$&$45.5^{46.1}_{44.8}$&$43.0^{43.6}_{42.3}$&$44.3^{44.9}_{43.7}$&$45.6^{46.2}_{
45.0}$&$43.3^{43.9}_{42.6}$&$44.6^{45.2}_{44.0}$&$45.9^{46.5}_{45.3}$&$43.6^{44.2}_{43.0}$&$45.0^{45.6}_{44.3}$&$46.3^{46.9}_{
45.6}$\\
14.75&$43.2^{43.8}_{42.5}$&$44.5^{45.1}_{43.9}$&$45.8^{46.3}_{45.2}$&$43.4^{43.9}_{42.7}$&$44.7^{45.3}_{44.0}$&$46.0^{46.5}_{
45.3}$&$43.6^{44.2}_{43.0}$&$45.0^{45.6}_{44.3}$&$46.3^{46.8}_{45.6}$&$44.0^{44.6}_{43.3}$&$45.3^{45.9}_{44.6}$&$46.6^{47.2}_{
46.0}$\\
15.00&$43.5^{44.1}_{42.9}$&$44.8^{45.3}_{44.2}$&$46.0^{46.5}_{45.4}$&$43.7^{44.2}_{43.0}$&$45.0^{45.5}_{44.3}$&$46.2^{46.8}_{
45.6}$&$44.0^{44.5}_{43.3}$&$45.3^{45.8}_{44.6}$&$46.5^{47.1}_{45.9}$&$44.3^{44.9}_{43.6}$&$45.6^{46.2}_{45.0}$&$46.9^{47.5}_{
46.3}$\\
15.25&$43.8^{44.3}_{43.2}$&$45.0^{45.5}_{44.4}$&$46.0^{46.5}_{45.5}$&$44.0^{44.5}_{43.3}$&$45.2^{45.8}_{44.6}$&$46.4^{46.9}_{
45.8}$&$44.3^{44.8}_{43.6}$&$45.5^{46.1}_{44.9}$&$46.8^{47.3}_{46.2}$&$44.6^{45.2}_{44.0}$&$45.9^{46.5}_{45.3}$&$47.2^{47.7}_{
46.5}$\\
15.50&$44.0^{44.5}_{43.4}$&$45.1^{45.5}_{44.5}$&$45.9^{46.2}_{45.5}$&$44.2^{44.8}_{43.6}$&$45.4^{45.9}_{44.8}$&$46.5^{47.0}_{
46.0}$&$44.5^{45.1}_{43.9}$&$45.8^{46.3}_{45.2}$&$47.0^{47.5}_{46.4}$&$44.9^{45.5}_{44.3}$&$46.2^{46.7}_{45.5}$&$47.4^{47.9}_{
46.8}$\\
15.75&$44.1^{44.5}_{43.6}$&$44.9^{45.2}_{44.5}$&$45.7^{46.1}_{45.3}$&$44.4^{45.0}_{43.8}$&$45.5^{46.0}_{45.0}$&$46.4^{46.8}_{
46.0}$&$44.8^{45.3}_{44.2}$&$46.0^{46.5}_{45.4}$&$47.0^{47.4}_{46.5}$&$45.2^{45.7}_{44.5}$&$46.4^{46.9}_{45.8}$&$47.5^{48.0}_{
47.0}$\\
16.00&$43.9^{44.3}_{43.5}$&$44.7^{45.1}_{44.3}$&$45.6^{46.0}_{45.1}$&$44.5^{45.0}_{44.0}$&$45.4^{45.8}_{45.0}$&$46.2^{46.6}_{
45.8}$&$45.0^{45.5}_{44.4}$&$46.0^{46.4}_{45.5}$&$46.8^{47.1}_{46.4}$&$45.4^{45.9}_{44.8}$&$46.5^{47.0}_{46.0}$&$47.5^{47.8}_{
47.0}$\\
16.25&$43.7^{44.1}_{43.3}$&$44.6^{45.0}_{44.2}$&$45.5^{45.9}_{45.0}$&$44.5^{44.8}_{44.0}$&$45.2^{45.6}_{44.8}$&$46.1^{46.5}_{
45.7}$&$45.0^{45.4}_{44.5}$&$45.8^{46.2}_{45.5}$&$46.6^{47.0}_{46.2}$&$45.5^{46.0}_{45.0}$&$46.5^{46.8}_{46.0}$&$47.2^{47.6}_{
46.9}$\\
16.50&$43.6^{44.0}_{43.2}$&$44.5^{44.9}_{44.1}$&$45.4^{45.8}_{44.9}$&$44.3^{44.6}_{43.9}$&$45.1^{45.5}_{44.7}$&$46.0^{46.4}_{
45.5}$&$44.9^{45.2}_{44.5}$&$45.6^{46.0}_{45.2}$&$46.5^{46.9}_{46.1}$&$45.5^{45.9}_{45.0}$&$46.2^{46.6}_{45.9}$&$47.0^{47.4}_{
46.6}$\\
16.75&$43.5^{43.9}_{43.1}$&$44.4^{44.8}_{44.0}$&$45.3^{45.7}_{44.8}$&$44.1^{44.5}_{43.7}$&$45.0^{45.4}_{44.6}$&$45.9^{46.2}_{
45.4}$&$44.7^{45.0}_{44.3}$&$45.5^{45.9}_{45.1}$&$46.4^{46.7}_{45.9}$&$45.3^{45.6}_{44.9}$&$46.1^{46.4}_{45.7}$&$46.9^{47.3}_{
46.5}$\\
17.00&$43.4^{43.8}_{43.0}$&$44.3^{44.7}_{43.8}$&$45.2^{45.6}_{44.7}$&$44.0^{44.4}_{43.6}$&$44.8^{45.2}_{44.4}$&$45.7^{46.1}_{
45.3}$&$44.5^{44.9}_{44.1}$&$45.3^{45.7}_{44.9}$&$46.2^{46.6}_{45.8}$&$45.1^{45.4}_{44.7}$&$45.9^{46.3}_{45.5}$&$46.7^{47.1}_{
46.3}$\\
17.25&$43.2^{43.7}_{42.8}$&$44.2^{44.6}_{43.7}$&$45.1^{45.5}_{44.6}$&$43.8^{44.1}_{43.3}$&$44.6^{45.0}_{44.2}$&$45.5^{45.9}_{
45.1}$&$44.2^{44.6}_{43.8}$&$45.1^{45.5}_{44.7}$&$45.9^{46.3}_{45.5}$&$44.8^{45.2}_{44.4}$&$45.6^{46.0}_{45.2}$&$46.4^{46.8}_{
46.0}$\\
17.50&$43.1^{43.5}_{42.6}$&$44.1^{44.5}_{43.6}$&$45.0^{45.5}_{44.5}$&$43.4^{43.8}_{43.0}$&$44.3^{44.8}_{43.9}$&$45.3^{45.7}_{
44.8}$&$43.8^{44.2}_{43.4}$&$44.7^{45.1}_{44.2}$&$45.5^{45.9}_{45.1}$&$44.3^{44.7}_{43.9}$&$45.1^{45.5}_{44.7}$&$45.9^{46.3}_{
45.5}$\\
17.75&$43.0^{43.5}_{42.6}$&$44.0^{44.5}_{43.5}$&$45.0^{45.5}_{44.5}$&$43.2^{43.7}_{42.8}$&$44.2^{44.6}_{43.7}$&$45.1^{45.6}_{
44.7}$&$43.4^{43.8}_{43.0}$&$44.3^{44.8}_{43.9}$&$45.3^{45.7}_{44.8}$&$43.7^{44.1}_{43.3}$&$44.6^{45.0}_{44.1}$&$45.4^{45.8}_{
45.0}$\\
18.00&$43.1^{43.5}_{42.6}$&$44.1^{44.5}_{43.6}$&$45.0^{45.5}_{44.6}$&$43.2^{43.6}_{42.7}$&$44.2^{44.6}_{43.7}$&$45.1^{45.5}_{
44.6}$&$43.3^{43.8}_{42.9}$&$44.3^{44.7}_{43.8}$&$45.2^{45.6}_{44.7}$&$43.6^{43.9}_{43.1}$&$44.4^{44.8}_{44.0}$&$45.3^{45.6}_{
44.8}$\\
18.25&$43.1^{43.6}_{42.6}$&$44.1^{44.5}_{43.6}$&$45.1^{45.5}_{44.6}$&$43.2^{43.6}_{42.7}$&$44.1^{44.6}_{43.7}$&$45.1^{45.5}_{
44.6}$&$43.3^{43.7}_{42.8}$&$44.2^{44.6}_{43.7}$&$45.1^{45.5}_{44.6}$&$43.4^{43.8}_{43.0}$&$44.3^{44.6}_{43.8}$&$45.1^{45.5}_{
44.7}$\\
18.50&$43.1^{43.6}_{42.6}$&$44.1^{44.6}_{43.6}$&$45.1^{45.5}_{44.6}$&$43.2^{43.6}_{42.7}$&$44.1^{44.6}_{43.7}$&$45.1^{45.5}_{
44.6}$&$43.2^{43.6}_{42.8}$&$44.1^{44.6}_{43.7}$&$45.0^{45.5}_{44.6}$&$43.3^{43.7}_{42.9}$&$44.2^{44.6}_{43.8}$&$45.0^{45.4}_{
44.6}$\\[1ex]
\hline

    \end{tabular}
    \label{tab:fig5}
\end{sidewaystable}

\FloatBarrier
\begin{sidewaystable}
\renewcommand{\arraystretch}{1.2}
\tablenum{A2}
\caption{Bolometric correction 
as a function of black hole mass (Figure \ref{fig:bc_mass})}
\tiny
\hspace{-0.7in}
\begin{tabular}{c@{\hspace{2em}}ccc@{\hspace{3em}}ccc@{\hspace{3em}}ccc@{\hspace{3em}}ccc}
    \hline\hline
    \textbf{spin = 0}
    &\multicolumn{3}{c}{$\log (\lambda_{Edd}) = -1.5$}
    &\multicolumn{3}{c}{$\log (\lambda_{Edd}) = -1.0$}
    &\multicolumn{3}{c}{$\log (\lambda_{Edd}) = -0.5$}
    &\multicolumn{3}{c}{$\log (\lambda_{Edd}) = 0.0$}\\
          \hline 
          $M_{\rm BH}=$
          &$10^7$--$10^8$
          &$10^8$--$10^9$
          &$10^9$--$10^{10}$
          &$10^7$--$10^8$
          &$10^8$--$10^9$
          &$10^9$--$10^{10}$
          &$10^7$--$10^8$
          &$10^8$--$10^9$
          &$10^9$--$10^{10}$
          &$10^7$--$10^8$
          &$10^8$--$10^9$
          &$10^9$--$10^{10}$\\
          \hline 
          $\log{\nu}$\\
14.00&$28.40^{42.72}_{18.37}$&$12.18^{17.60}_{8.382}$&$6.097^{8.096}_{4.834}$&$57.87^{87.66}_{36.83}$&$23.69^{35.22}_{15.41}$&
$10.18^{14.78}_{6.841}$&$81.72^{123.3}_{52.19}$&$33.64^{49.92}_{21.92}$&$14.44^{21.01}_{9.624}$&$112.0^{168.1}_{71.90}$&$46.61
^{68.82}_{30.51}$&$20.17^{29.25}_{13.48}$\\
14.25&$28.40^{42.72}_{18.37}$&$12.18^{17.60}_{8.382}$&$6.097^{8.096}_{4.834}$&$57.87^{87.66}_{36.83}$&$23.69^{35.22}_{15.41}$&
$10.18^{14.78}_{6.841}$&$81.72^{123.3}_{52.19}$&$33.64^{49.92}_{21.92}$&$14.44^{21.01}_{9.624}$&$112.0^{168.1}_{71.90}$&$46.61
^{68.82}_{30.51}$&$20.17^{29.25}_{13.48}$\\
14.50&$17.61^{25.95}_{11.72}$&$8.051^{11.26}_{5.797}$&$4.470^{5.628}_{3.810}$&$36.40^{54.17}_{23.68}$&$15.61^{22.70}_{10.45}$&
$7.130^{10.05}_{4.985}$&$52.15^{77.30}_{34.01}$&$22.43^{32.61}_{14.98}$&$10.14^{14.40}_{6.966}$&$72.61^{107.1}_{47.60}$&$31.53
^{45.66}_{21.11}$&$14.29^{20.29}_{9.789}$\\
14.75&$6.344^{8.596}_{4.738}$&$3.775^{4.615}_{3.285}$&$3.236^{3.505}_{3.183}$&$13.22^{18.57}_{9.219}$&$6.569^{8.901}_{4.811}$&
$3.658^{4.672}_{2.930}$&$19.54^{27.51}_{13.53}$&$9.489^{13.05}_{6.766}$&$4.925^{6.548}_{3.686}$&$28.18^{39.66}_{19.45}$&$13.56
^{18.75}_{9.552}$&$6.819^{9.229}_{4.951}$\\
15.00&$3.703^{4.401}_{3.330}$&$3.379^{3.807}_{3.290}$&$4.729^{5.407}_{3.876}$&$6.567^{8.744}_{4.908}$&$3.811^{4.776}_{3.128}$&
$2.776^{3.078}_{2.713}$&$9.530^{12.94}_{6.888}$&$5.084^{6.674}_{3.868}$&$3.073^{3.771}_{2.601}$&$13.65^{18.72}_{9.704}$&$6.991
^{9.384}_{5.131}$&$3.869^{4.982}_{3.031}$\\
15.25&$3.569^{4.103}_{3.429}$&$5.267^{6.228}_{4.191}$&$6.846^{7.035}_{6.312}$&$3.906^{4.864}_{3.236}$&$2.920^{3.188}_{2.869}$&
$3.421^{4.315}_{2.941}$&$5.182^{6.774}_{3.967}$&$3.180^{3.871}_{2.725}$&$2.609^{2.744}_{2.566}$&$7.098^{9.501}_{5.233}$&$3.969
^{5.083}_{3.137}$&$2.632^{3.073}_{2.407}$\\
15.50&$5.493^{6.631}_{4.320}$&$7.562^{8.052}_{6.743}$&$8.527^{8.785}_{8.101}$&$2.980^{3.235}_{2.930}$&$3.541^{4.534}_{3.019}$&
$6.240^{7.355}_{4.674}$&$3.231^{3.920}_{2.782}$&$2.688^{2.846}_{2.637}$&$3.695^{5.118}_{2.892}$&$4.022^{5.139}_{3.190}$&$2.689
^{3.126}_{2.476}$&$2.610^{3.150}_{2.469}$\\
15.75&$7.835^{8.552}_{6.878}$&$9.414^{10.05}_{8.634}$&$10.64^{10.89}_{10.12}$&$3.604^{4.647}_{3.042}$&$6.439^{7.699}_{4.796}$&
$8.714^{9.410}_{7.810}$&$2.705^{2.901}_{2.656}$&$3.822^{5.343}_{2.952}$&$7.843^{9.418}_{5.556}$&$2.691^{3.110}_{2.499}$&$2.669
^{3.261}_{2.495}$&$5.006^{7.558}_{3.367}$\\
16.00&$9.627^{10.55}_{8.639}$&$11.60^{12.29}_{10.65}$&$12.86^{13.02}_{12.36}$&$6.301^{7.617}_{4.683}$&$8.821^{9.757}_{7.740}$&
$10.95^{11.89}_{9.864}$&$3.777^{5.264}_{2.926}$&$7.778^{9.506}_{5.475}$&$11.06^{12.26}_{9.664}$&$2.659^{3.244}_{2.492}$&$4.966
^{7.511}_{3.343}$&$10.90^{12.82}_{7.838}$\\
16.25&$11.81^{12.83}_{10.62}$&$13.92^{14.58}_{12.94}$&$15.00^{15.04}_{14.64}$&$8.612^{9.685}_{7.437}$&$11.13^{12.38}_{9.811}$&
$13.88^{15.00}_{12.52}$&$7.487^{9.259}_{5.266}$&$11.00^{12.46}_{9.426}$&$14.41^{16.03}_{12.63}$&$4.818^{7.260}_{3.271}$&$10.63
^{12.75}_{7.578}$&$15.08^{17.12}_{12.94}$\\
16.50&$14.14^{15.15}_{12.89}$&$16.14^{16.65}_{15.25}$&$16.77^{16.83}_{16.55}$&$10.98^{12.42}_{9.518}$&$14.26^{15.76}_{12.58}$&
$17.52^{18.79}_{15.93}$&$10.66^{12.29}_{8.981}$&$14.51^{16.50}_{12.48}$&$19.08^{21.18}_{16.73}$&$10.13^{12.32}_{7.164}$&$14.88
^{17.26}_{12.56}$&$20.46^{23.25}_{17.51}$\\
16.75&$16.46^{17.32}_{15.30}$&$18.05^{18.30}_{17.41}$&$18.14^{18.31}_{17.51}$&$14.50^{16.28}_{12.61}$&$18.49^{20.26}_{16.48}$&
$22.26^{23.65}_{20.45}$&$14.61^{16.90}_{12.35}$&$19.93^{22.60}_{17.16}$&$26.01^{28.76}_{22.90}$&$14.75^{17.36}_{12.18}$&$21.01
^{24.40}_{17.70}$&$28.91^{32.83}_{24.74}$\\
17.00&$18.52^{19.10}_{17.62}$&$19.39^{19.45}_{19.15}$&$18.79^{19.31}_{17.81}$&$19.64^{21.77}_{17.30}$&$24.37^{26.36}_{22.01}$&
$28.52^{29.93}_{26.58}$&$21.25^{24.46}_{18.03}$&$28.67^{32.33}_{24.82}$&$36.98^{40.65}_{32.75}$&$22.20^{26.13}_{18.31}$&$31.59
^{36.64}_{26.65}$&$43.36^{49.17}_{37.15}$\\
17.25&$20.42^{20.56}_{19.99}$&$20.40^{20.56}_{19.86}$&$18.91^{19.78}_{17.59}$&$29.07^{31.53}_{26.25}$&$34.38^{36.41}_{31.79}$&
$38.41^{39.56}_{36.64}$&$36.41^{41.44}_{31.31}$&$47.87^{53.36}_{41.99}$&$60.19^{65.33}_{53.99}$&$41.22^{48.31}_{34.16}$&$58.12
^{67.23}_{49.27}$&$79.24^{89.49}_{68.08}$\\
17.50&$21.12^{21.23}_{20.80}$&$20.20^{20.74}_{19.30}$&$18.08^{19.19}_{16.58}$&$45.95^{47.96}_{43.43}$&$50.14^{51.38}_{48.14}$&
$52.32^{52.50}_{51.55}$&$81.07^{89.39}_{72.53}$&$99.41^{107.6}_{90.22}$&$117.3^{123.7}_{108.5}$&$120.1^{138.4}_{101.7}$&$163.4
^{186.9}_{141.1}$&$216.9^{241.5}_{188.6}$\\
17.75&$18.94^{19.19}_{18.52}$&$17.88^{18.45}_{16.98}$&$15.83^{16.88}_{14.45}$&$55.42^{56.63}_{53.81}$&$57.90^{58.45}_{56.72}$&
$58.52^{58.74}_{57.89}$&$137.8^{147.3}_{128.7}$&$158.1^{166.8}_{148.1}$&$176.3^{182.8}_{167.8}$&$323.8^{360.4}_{285.7}$&$411.9
^{462.2}_{368.1}$&$522.9^{567.9}_{462.9}$\\
18.00&$16.19^{16.41}_{15.83}$&$15.27^{15.77}_{14.50}$&$13.52^{14.41}_{12.34}$&$55.75^{56.92}_{54.20}$&$58.11^{58.67}_{56.99}$&
$58.71^{58.93}_{58.02}$&$157.8^{168.5}_{147.8}$&$180.5^{190.1}_{169.3}$&$200.5^{207.9}_{191.3}$&$433.0^{480.9}_{383.0}$&$547.6
^{613.7}_{490.5}$&$692.7^{750.1}_{613.7}$\\
18.25&$13.77^{13.96}_{13.47}$&$12.99^{13.41}_{12.34}$&$11.49^{12.26}_{10.49}$&$55.80^{56.95}_{54.25}$&$58.14^{58.72}_{57.03}$&
$58.73^{58.98}_{58.04}$&$178.2^{190.3}_{167.0}$&$203.9^{214.7}_{191.2}$&$226.3^{234.9}_{216.1}$&$556.6^{618.8}_{491.7}$&$704.4
^{789.2}_{629.8}$&$890.5^{963.2}_{788.1}$\\
18.50&$12.66^{12.83}_{12.37}$&$11.94^{12.32}_{11.33}$&$10.56^{11.26}_{9.633}$&$55.92^{57.07}_{54.37}$&$58.26^{58.84}_{57.14}$&
$58.85^{59.10}_{58.15}$&$190.4^{203.4}_{178.5}$&$217.9^{229.5}_{204.3}$&$241.8^{251.0}_{230.9}$&$637.0^{708.5}_{562.2}$&$806.5
^{903.4}_{720.4}$&$1019.^{1102.}_{901.6}$\\[0.5ex]
\hline
\textbf{spin = 1}\\
14.00&$99.93^{150.8}_{63.70}$&$41.00^{60.93}_{26.66}$&$17.53^{25.56}_{11.68}$&$169.2^{250.0}_{110.6}$&$72.98^{106.0}_{48.62}$&
$32.72^{46.71}_{22.22}$&$231.4^{339.1}_{152.6}$&$101.6^{146.4}_{68.28}$&$46.29^{65.65}_{31.63}$&$320.7^{467.1}_{213.1}$&$142.8
^{204.7}_{96.45}$&$65.68^{92.77}_{45.02}$\\
14.25&$99.93^{150.8}_{63.70}$&$41.00^{60.93}_{26.66}$&$17.53^{25.56}_{11.68}$&$169.2^{250.0}_{110.6}$&$72.98^{106.0}_{48.62}$&
$32.72^{46.71}_{22.22}$&$231.4^{339.1}_{152.6}$&$101.6^{146.4}_{68.28}$&$46.29^{65.65}_{31.63}$&$320.7^{467.1}_{213.1}$&$142.8
^{204.7}_{96.45}$&$65.68^{92.77}_{45.02}$\\
14.50&$63.89^{94.79}_{41.55}$&$27.34^{39.83}_{18.20}$&$12.29^{17.50}_{8.431}$&$113.8^{165.2}_{75.91}$&$51.11^{72.95}_{34.73}$&
$23.82^{33.43}_{16.48}$&$159.1^{229.2}_{106.9}$&$72.52^{102.8}_{49.56}$&$34.12^{47.73}_{23.66}$&$224.4^{321.6}_{151.7}$&$103.3
^{145.9}_{70.74}$&$48.77^{68.13}_{33.79}$\\
14.75&$23.92^{33.81}_{16.44}$&$11.46^{15.85}_{8.124}$&$5.896^{7.863}_{4.419}$&$47.60^{66.51}_{33.02}$&$23.06^{31.86}_{16.21}$&
$11.50^{15.65}_{8.260}$&$68.94^{96.20}_{47.82}$&$33.33^{46.12}_{23.35}$&$16.46^{22.54}_{11.70}$&$99.42^{138.8}_{68.85}$&$47.84
^{66.38}_{33.37}$&$23.40^{32.20}_{16.50}$\\
15.00&$11.47^{15.68}_{8.212}$&$6.026^{7.953}_{4.577}$&$3.659^{4.466}_{3.145}$&$23.20^{31.92}_{16.38}$&$11.67^{15.83}_{8.428}$&
$6.205^{8.165}_{4.697}$&$33.58^{46.36}_{23.58}$&$16.67^{22.77}_{11.90}$&$8.589^{11.51}_{6.308}$&$48.19^{66.78}_{33.69}$&$23.66
^{32.51}_{16.73}$&$11.93^{16.16}_{8.607}$\\
15.25&$6.100^{8.024}_{4.651}$&$3.742^{4.539}_{3.248}$&$3.173^{3.424}_{3.121}$&$11.76^{15.92}_{8.513}$&$6.289^{8.254}_{4.782}$&
$3.797^{4.662}_{3.208}$&$16.80^{22.92}_{12.01}$&$8.692^{11.62}_{6.405}$&$4.844^{6.221}_{3.803}$&$23.84^{32.72}_{16.88}$&$12.06
^{16.31}_{8.719}$&$6.415^{8.447}_{4.833}$\\
15.50&$3.769^{4.564}_{3.279}$&$3.217^{3.494}_{3.161}$&$4.489^{5.745}_{3.553}$&$6.323^{8.289}_{4.817}$&$3.833^{4.699}_{3.249}$&
$3.056^{3.210}_{3.005}$&$8.747^{11.68}_{6.455}$&$4.890^{6.270}_{3.849}$&$3.203^{3.768}_{2.884}$&$12.14^{16.41}_{8.786}$&$6.471
^{8.514}_{4.885}$&$3.824^{4.758}_{3.150}$\\
15.75&$3.199^{3.509}_{3.145}$&$4.537^{5.830}_{3.573}$&$7.310^{8.166}_{5.977}$&$3.785^{4.625}_{3.223}$&$3.052^{3.186}_{3.004}$&
$3.871^{4.951}_{3.196}$&$4.824^{6.172}_{3.808}$&$3.183^{3.729}_{2.886}$&$2.943^{3.376}_{2.856}$&$6.378^{8.380}_{4.826}$&$3.788
^{4.702}_{3.134}$&$2.802^{3.087}_{2.751}$\\
16.00&$4.397^{5.639}_{3.484}$&$7.125^{8.077}_{5.785}$&$9.189^{10.15}_{8.174}$&$3.019^{3.157}_{2.964}$&$3.784^{4.826}_{3.137}$&
$6.358^{7.381}_{4.965}$&$3.168^{3.718}_{2.863}$&$2.909^{3.325}_{2.830}$&$4.453^{5.823}_{3.396}$&$3.783^{4.702}_{3.128}$&$2.788
^{3.080}_{2.732}$&$3.189^{4.050}_{2.776}$\\
16.25&$6.755^{7.745}_{5.458}$&$8.955^{10.06}_{7.851}$&$11.49^{12.68}_{10.18}$&$3.629^{4.593}_{3.038}$&$6.057^{7.109}_{4.717}$&
$8.279^{9.329}_{7.216}$&$2.838^{3.216}_{2.772}$&$4.268^{5.567}_{3.281}$&$7.142^{8.234}_{5.719}$&$2.747^{3.047}_{2.684}$&$3.098
^{3.905}_{2.722}$&$5.577^{7.122}_{4.027}$\\
16.50&$8.537^{9.713}_{7.375}$&$11.28^{12.64}_{9.850}$&$14.30^{15.62}_{12.78}$&$5.631^{6.687}_{4.383}$&$7.939^{9.088}_{6.797}$&
$10.69^{12.12}_{9.226}$&$4.000^{5.191}_{3.116}$&$6.732^{7.886}_{5.335}$&$9.407^{10.85}_{8.017}$&$2.961^{3.692}_{2.625}$&$5.223
^{6.711}_{3.803}$&$8.368^{9.770}_{6.871}$\\
16.75&$11.06^{12.52}_{9.525}$&$14.37^{15.90}_{12.69}$&$17.71^{19.09}_{16.06}$&$7.597^{8.857}_{6.359}$&$10.62^{12.23}_{9.006}$&
$14.35^{16.20}_{12.41}$&$6.297^{7.531}_{4.930}$&$9.207^{10.80}_{7.674}$&$13.05^{15.10}_{11.00}$&$4.857^{6.277}_{3.549}$&$8.015
^{9.549}_{6.436}$&$11.73^{13.84}_{9.737}$\\
17.00&$14.65^{16.32}_{12.82}$&$18.35^{19.96}_{16.51}$&$21.79^{23.08}_{20.13}$&$10.79^{12.61}_{8.955}$&$15.04^{17.18}_{12.82}$&
$19.95^{22.32}_{17.42}$&$9.165^{11.01}_{7.399}$&$13.59^{15.97}_{11.23}$&$19.25^{22.21}_{16.26}$&$7.714^{9.467}_{5.976}$&$12.01
^{14.44}_{9.682}$&$17.84^{21.07}_{14.74}$\\
17.25&$20.74^{22.47}_{18.74}$&$24.43^{25.88}_{22.65}$&$27.39^{28.34}_{26.04}$&$18.18^{20.90}_{15.33}$&$24.38^{27.36}_{21.22}$&
$31.10^{34.19}_{27.68}$&$16.67^{19.95}_{13.41}$&$24.42^{28.46}_{20.35}$&$34.00^{38.93}_{28.96}$&$14.47^{17.88}_{11.25}$&$22.77
^{27.37}_{18.30}$&$33.72^{39.80}_{27.94}$\\
17.50&$29.86^{30.96}_{28.47}$&$32.11^{32.84}_{31.08}$&$33.44^{33.63}_{32.92}$&$38.56^{42.29}_{34.45}$&$46.77^{50.40}_{42.79}$&
$54.69^{58.05}_{50.74}$&$46.19^{53.61}_{38.55}$&$63.26^{71.55}_{54.55}$&$83.01^{92.83}_{72.71}$&$47.35^{57.69}_{37.54}$&$72.51
^{86.45}_{59.00}$&$105.1^{123.7}_{88.24}$\\
17.75&$32.43^{32.93}_{31.78}$&$33.44^{33.69}_{32.98}$&$33.76^{33.82}_{33.62}$&$59.58^{62.70}_{56.38}$&$66.36^{69.37}_{63.27}$&
$72.90^{75.66}_{69.57}$&$108.7^{119.9}_{97.11}$&$134.0^{145.7}_{121.5}$&$162.2^{176.6}_{147.5}$&$176.8^{204.6}_{150.0}$&$248.7
^{290.5}_{208.8}$&$341.8^{400.3}_{292.5}$\\
18.00&$30.20^{30.63}_{29.61}$&$31.08^{31.31}_{30.68}$&$31.36^{31.41}_{31.21}$&$63.48^{66.68}_{60.29}$&$70.41^{73.52}_{67.31}$&
$77.16^{80.03}_{73.70}$&$134.3^{147.6}_{120.4}$&$164.2^{178.4}_{149.2}$&$197.3^{215.1}_{180.3}$&$264.3^{301.5}_{223.8}$&$366.1
^{427.2}_{307.9}$&$499.6^{586.5}_{427.2}$\\
18.25&$27.95^{28.35}_{27.41}$&$28.77^{28.98}_{28.40}$&$29.02^{29.07}_{28.88}$&$66.73^{70.12}_{63.43}$&$74.04^{77.31}_{70.82}$&
$81.16^{84.19}_{77.48}$&$160.4^{176.6}_{143.8}$&$196.4^{213.5}_{178.1}$&$235.6^{257.3}_{215.6}$&$366.2^{418.1}_{308.9}$&$507.4
^{593.2}_{425.8}$&$691.9^{814.6}_{590.1}$\\
18.50&$26.87^{27.25}_{26.35}$&$27.65^{27.85}_{27.29}$&$27.90^{27.94}_{27.76}$&$68.65^{72.14}_{65.27}$&$76.17^{79.54}_{72.89}$&
$83.50^{86.63}_{79.70}$&$176.7^{194.5}_{158.2}$&$216.4^{235.3}_{196.1}$&$259.4^{283.5}_{237.4}$&$435.4^{498.5}_{366.9}$&$604.2
^{707.1}_{506.3}$&$823.7^{971.4}_{701.4}$\\
\hline

    \end{tabular}
    \label{tab:fig6}
\end{sidewaystable}

\FloatBarrier
\begin{sidewaystable}
\renewcommand{\arraystretch}{1.2}
\tablenum{A3}
\caption{Accretion disk SED 
as a function of Eddington ratio (Figure \ref{fig:ad_mdot})}
\tiny
    \begin{tabular}{c@{\hspace{2em}}ccc@{\hspace{3em}}ccc@{\hspace{3em}}ccc@{\hspace{3em}}ccc}
         \hline\hline
         \textbf{spin = 0}
         &\multicolumn{3}{c}{$M_{\rm BH}/M_{\odot} = 10^7$}
         &\multicolumn{3}{c}{$M_{\rm BH}/M_{\odot} = 10^8$}
         &\multicolumn{3}{c}{$M_{\rm BH}/M_{\odot} = 10^9$}
         &\multicolumn{3}{c}{$M_{\rm BH}/M_{\odot} = 10^{10}$}\\[0.5ex]
          \hline 
          $\lambda_{\rm Edd}=$
          &$-1.5$--$-1.0$
          &$-1.0$--$-0.5$
          &$-0.5$--$0.0$
          &$-1.5$--$-1.0$
          &$-1.0$--$-0.5$
          &$-0.5$--$0.0$
          &$-1.5$--$-1.0$
          &$-1.0$--$-0.5$
          &$-0.5$--$0.0$
          &$-1.5$--$-1.0$
          &$-1.0$--$-0.5$
          &$-0.5$--$0.0$\\[0.5ex]
          \hline
$\log(\nu)$\\
14.00&$42.2^{42.4}_{42.1}$&$42.6^{42.8}_{42.4}$&$43.0^{43.2}_{42.8}$&$43.6^{43.8}_{43.4}$&$44.0^{44.2}_{43.8}$&$44.4^{44.5}_{
44.2}$&$45.0^{45.1}_{44.7}$&$45.4^{45.5}_{45.2}$&$45.7^{45.9}_{45.6}$&$46.3^{46.5}_{45.9}$&$46.7^{46.9}_{46.5}$&$47.1^{47.3}_{
46.9}$\\
14.25&$42.2^{42.4}_{42.1}$&$42.6^{42.8}_{42.4}$&$43.0^{43.2}_{42.8}$&$43.6^{43.8}_{43.4}$&$44.0^{44.2}_{43.8}$&$44.4^{44.5}_{
44.2}$&$45.0^{45.1}_{44.7}$&$45.4^{45.5}_{45.2}$&$45.7^{45.9}_{45.6}$&$46.3^{46.5}_{45.9}$&$46.7^{46.9}_{46.5}$&$47.1^{47.3}_{
46.9}$\\
14.50&$42.4^{42.6}_{42.3}$&$42.8^{43.0}_{42.6}$&$43.2^{43.4}_{43.0}$&$43.8^{44.0}_{43.6}$&$44.2^{44.4}_{44.0}$&$44.6^{44.7}_{
44.4}$&$45.1^{45.3}_{44.9}$&$45.5^{45.7}_{45.3}$&$45.9^{46.1}_{45.7}$&$46.4^{46.6}_{46.0}$&$46.8^{47.0}_{46.6}$&$47.2^{47.4}_{
47.0}$\\
14.75&$42.9^{43.1}_{42.8}$&$43.3^{43.4}_{43.1}$&$43.6^{43.8}_{43.5}$&$44.2^{44.4}_{44.0}$&$44.6^{44.8}_{44.4}$&$45.0^{45.1}_{
44.8}$&$45.4^{45.6}_{45.1}$&$45.9^{46.0}_{45.7}$&$46.2^{46.4}_{46.1}$&$46.5^{46.8}_{46.0}$&$47.1^{47.3}_{46.8}$&$47.5^{47.7}_{
47.3}$\\
15.00&$43.2^{43.4}_{43.0}$&$43.6^{43.8}_{43.4}$&$44.0^{44.1}_{43.8}$&$44.5^{44.7}_{44.1}$&$44.9^{45.0}_{44.7}$&$45.2^{45.4}_{
45.1}$&$45.6^{45.8}_{45.0}$&$46.1^{46.3}_{45.8}$&$46.5^{46.7}_{46.3}$&$46.4^{46.8}_{45.8}$&$47.2^{47.4}_{46.9}$&$47.7^{47.9}_{
47.5}$\\
15.25&$43.5^{43.7}_{43.2}$&$43.9^{44.0}_{43.7}$&$44.2^{44.4}_{44.1}$&$44.6^{44.8}_{44.0}$&$45.1^{45.3}_{44.9}$&$45.5^{45.7}_{
45.3}$&$45.4^{45.8}_{44.8}$&$46.2^{46.4}_{45.9}$&$46.7^{46.9}_{46.5}$&$46.2^{46.6}_{45.7}$&$47.1^{47.4}_{46.7}$&$47.7^{48.0}_{
47.4}$\\
15.50&$43.6^{43.8}_{43.1}$&$44.1^{44.3}_{43.9}$&$44.5^{44.7}_{44.3}$&$44.5^{44.8}_{43.8}$&$45.2^{45.4}_{44.9}$&$45.7^{45.9}_{
45.5}$&$45.2^{45.6}_{44.7}$&$46.1^{46.4}_{45.7}$&$46.7^{47.0}_{46.4}$&$46.1^{46.4}_{45.6}$&$46.8^{47.1}_{46.4}$&$47.5^{47.9}_{
47.1}$\\
15.75&$43.5^{43.8}_{42.8}$&$44.2^{44.4}_{43.9}$&$44.7^{44.9}_{44.5}$&$44.2^{44.6}_{43.7}$&$45.1^{45.4}_{44.7}$&$45.7^{46.0}_{
45.4}$&$45.1^{45.4}_{44.6}$&$45.8^{46.1}_{45.4}$&$46.5^{46.9}_{46.2}$&$46.0^{46.3}_{45.5}$&$46.6^{46.9}_{46.3}$&$47.2^{47.5}_{
46.9}$\\
16.00&$43.3^{43.6}_{42.8}$&$44.1^{44.4}_{43.7}$&$44.7^{45.0}_{44.4}$&$44.1^{44.4}_{43.6}$&$44.8^{45.1}_{44.5}$&$45.5^{45.9}_{
45.2}$&$45.0^{45.3}_{44.5}$&$45.6^{45.9}_{45.3}$&$46.2^{46.5}_{45.9}$&$45.9^{46.2}_{45.4}$&$46.5^{46.7}_{46.2}$&$47.0^{47.2}_{
46.8}$\\
16.25&$43.1^{43.5}_{42.7}$&$43.8^{44.2}_{43.5}$&$44.6^{44.9}_{44.2}$&$44.0^{44.3}_{43.6}$&$44.6^{44.9}_{44.4}$&$45.2^{45.5}_{
44.9}$&$44.9^{45.2}_{44.5}$&$45.5^{45.8}_{45.2}$&$46.0^{46.3}_{45.8}$&$45.8^{46.1}_{45.4}$&$46.4^{46.6}_{46.1}$&$46.9^{47.1}_{
46.7}$\\
16.50&$43.0^{43.4}_{42.6}$&$43.7^{43.9}_{43.4}$&$44.2^{44.5}_{43.9}$&$43.9^{44.2}_{43.5}$&$44.5^{44.8}_{44.3}$&$45.1^{45.3}_{
44.8}$&$44.8^{45.1}_{44.4}$&$45.4^{45.6}_{45.1}$&$45.9^{46.1}_{45.7}$&$45.7^{46.0}_{45.3}$&$46.3^{46.5}_{46.0}$&$46.8^{47.0}_{
46.5}$\\
16.75&$42.9^{43.2}_{42.5}$&$43.5^{43.8}_{43.3}$&$44.1^{44.3}_{43.8}$&$43.8^{44.1}_{43.4}$&$44.4^{44.6}_{44.1}$&$44.9^{45.1}_{
44.7}$&$44.7^{45.0}_{44.4}$&$45.3^{45.5}_{45.0}$&$45.8^{46.0}_{45.5}$&$45.7^{45.9}_{45.3}$&$46.2^{46.4}_{45.9}$&$46.6^{46.8}_{
46.4}$\\
17.00&$42.8^{43.1}_{42.4}$&$43.4^{43.6}_{43.1}$&$43.9^{44.1}_{43.6}$&$43.7^{44.0}_{43.4}$&$44.3^{44.5}_{44.0}$&$44.7^{45.0}_{
44.5}$&$44.7^{44.9}_{44.3}$&$45.1^{45.3}_{44.9}$&$45.6^{45.8}_{45.4}$&$45.6^{45.8}_{45.3}$&$46.0^{46.2}_{45.8}$&$46.5^{46.7}_{
46.2}$\\
17.25&$42.7^{42.9}_{42.4}$&$43.2^{43.4}_{42.9}$&$43.6^{43.9}_{43.4}$&$43.6^{43.8}_{43.4}$&$44.1^{44.2}_{43.9}$&$44.5^{44.7}_{
44.3}$&$44.6^{44.8}_{44.3}$&$45.0^{45.1}_{44.8}$&$45.3^{45.5}_{45.1}$&$45.5^{45.7}_{45.3}$&$45.9^{46.0}_{45.7}$&$46.2^{46.4}_{
46.0}$\\
17.50&$42.6^{42.7}_{42.4}$&$42.9^{43.0}_{42.7}$&$43.2^{43.4}_{43.0}$&$43.5^{43.7}_{43.3}$&$43.8^{43.9}_{43.7}$&$44.1^{44.2}_{
43.9}$&$44.5^{44.6}_{44.3}$&$44.7^{44.8}_{44.6}$&$45.0^{45.1}_{44.8}$&$45.5^{45.6}_{45.3}$&$45.7^{45.7}_{45.6}$&$45.9^{46.0}_{
45.7}$\\
17.75&$42.5^{42.6}_{42.4}$&$42.7^{42.8}_{42.6}$&$42.8^{42.9}_{42.8}$&$43.5^{43.6}_{43.4}$&$43.7^{43.7}_{43.6}$&$43.8^{43.8}_{
43.7}$&$44.5^{44.6}_{44.4}$&$44.6^{44.6}_{44.6}$&$44.7^{44.7}_{44.6}$&$45.5^{45.5}_{45.4}$&$45.6^{45.6}_{45.5}$&$45.6^{45.6}_{
45.6}$\\
18.00&$42.6^{42.6}_{42.5}$&$42.7^{42.7}_{42.6}$&$42.8^{42.8}_{42.7}$&$43.6^{43.6}_{43.5}$&$43.6^{43.6}_{43.6}$&$43.7^{43.7}_{
43.6}$&$44.5^{44.6}_{44.5}$&$44.6^{44.6}_{44.6}$&$44.6^{44.6}_{44.6}$&$45.5^{45.5}_{45.5}$&$45.5^{45.5}_{45.5}$&$45.5^{45.5}_{
45.5}$\\
18.25&$42.6^{42.6}_{42.5}$&$42.6^{42.7}_{42.6}$&$42.7^{42.7}_{42.7}$&$43.6^{43.6}_{43.5}$&$43.6^{43.6}_{43.6}$&$43.6^{43.6}_{
43.6}$&$44.6^{44.6}_{44.5}$&$44.6^{44.6}_{44.5}$&$44.5^{44.5}_{44.5}$&$45.6^{45.6}_{45.5}$&$45.5^{45.5}_{45.5}$&$45.4^{45.5}_{
45.4}$\\
18.50&$42.6^{42.6}_{42.6}$&$42.6^{42.6}_{42.6}$&$42.6^{42.6}_{42.6}$&$43.6^{43.6}_{43.6}$&$43.6^{43.6}_{43.6}$&$43.5^{43.6}_{
43.5}$&$44.6^{44.6}_{44.6}$&$44.5^{44.6}_{44.5}$&$44.5^{44.5}_{44.4}$&$45.6^{45.6}_{45.5}$&$45.5^{45.5}_{45.4}$&$45.4^{45.4}_{
45.3}$\\[0.5ex]
         \hline
         \textbf{spin = 1}\\
14.00&$42.0^{42.1}_{41.9}$&$42.3^{42.4}_{42.1}$&$42.6^{42.8}_{42.5}$&$43.3^{43.5}_{43.3}$&$43.7^{43.8}_{43.5}$&$44.0^{44.2}_{
43.8}$&$44.7^{44.8}_{44.6}$&$45.0^{45.2}_{44.8}$&$45.3^{45.5}_{45.2}$&$46.0^{46.2}_{46.0}$&$46.3^{46.5}_{46.2}$&$46.7^{46.8}_{
46.5}$\\
14.25&$42.0^{42.1}_{41.9}$&$42.3^{42.4}_{42.1}$&$42.6^{42.8}_{42.5}$&$43.3^{43.5}_{43.3}$&$43.7^{43.8}_{43.5}$&$44.0^{44.2}_{
43.8}$&$44.7^{44.8}_{44.6}$&$45.0^{45.2}_{44.8}$&$45.3^{45.5}_{45.2}$&$46.0^{46.2}_{46.0}$&$46.3^{46.5}_{46.2}$&$46.7^{46.8}_{
46.5}$\\
14.50&$42.2^{42.3}_{42.1}$&$42.5^{42.6}_{42.3}$&$42.8^{43.0}_{42.6}$&$43.5^{43.6}_{43.5}$&$43.8^{44.0}_{43.7}$&$44.1^{44.3}_{
44.0}$&$44.9^{45.0}_{44.8}$&$45.1^{45.3}_{45.0}$&$45.5^{45.6}_{45.3}$&$46.2^{46.3}_{46.1}$&$46.5^{46.6}_{46.3}$&$46.8^{46.9}_{
46.6}$\\
14.75&$42.6^{42.7}_{42.5}$&$42.8^{43.0}_{42.7}$&$43.2^{43.3}_{43.0}$&$43.9^{44.0}_{43.9}$&$44.2^{44.3}_{44.0}$&$44.5^{44.6}_{
44.3}$&$45.2^{45.3}_{45.2}$&$45.5^{45.6}_{45.3}$&$45.8^{46.0}_{45.6}$&$46.5^{46.6}_{46.4}$&$46.8^{46.9}_{46.6}$&$47.1^{47.3}_{
46.9}$\\
15.00&$42.9^{43.0}_{42.9}$&$43.2^{43.3}_{43.0}$&$43.5^{43.6}_{43.3}$&$44.2^{44.3}_{44.2}$&$44.5^{44.6}_{44.3}$&$44.8^{45.0}_{
44.6}$&$45.5^{45.6}_{45.4}$&$45.8^{45.9}_{45.6}$&$46.1^{46.3}_{45.9}$&$46.7^{46.8}_{46.5}$&$47.0^{47.2}_{46.8}$&$47.4^{47.5}_{
47.2}$\\
15.25&$43.2^{43.3}_{43.2}$&$43.5^{43.6}_{43.3}$&$43.8^{44.0}_{43.6}$&$44.5^{44.6}_{44.4}$&$44.8^{44.9}_{44.6}$&$45.1^{45.3}_{
44.9}$&$45.7^{45.8}_{45.5}$&$46.0^{46.2}_{45.8}$&$46.4^{46.5}_{46.2}$&$46.8^{47.0}_{46.5}$&$47.2^{47.4}_{47.0}$&$47.6^{47.8}_{
47.4}$\\
15.50&$43.5^{43.6}_{43.4}$&$43.8^{43.9}_{43.6}$&$44.1^{44.3}_{43.9}$&$44.7^{44.8}_{44.5}$&$45.0^{45.2}_{44.8}$&$45.4^{45.5}_{
45.2}$&$45.8^{46.0}_{45.5}$&$46.2^{46.4}_{46.0}$&$46.6^{46.8}_{46.4}$&$46.7^{47.0}_{46.3}$&$47.3^{47.5}_{47.0}$&$47.8^{48.0}_{
47.5}$\\
15.75&$43.7^{43.8}_{43.6}$&$44.0^{44.2}_{43.8}$&$44.4^{44.5}_{44.2}$&$44.8^{45.0}_{44.5}$&$45.2^{45.4}_{45.0}$&$45.6^{45.8}_{
45.4}$&$45.7^{46.0}_{45.3}$&$46.3^{46.5}_{46.0}$&$46.8^{47.0}_{46.5}$&$46.5^{46.8}_{46.1}$&$47.1^{47.4}_{46.8}$&$47.8^{48.0}_{
47.5}$\\
16.00&$43.8^{44.0}_{43.5}$&$44.2^{44.4}_{44.0}$&$44.6^{44.8}_{44.4}$&$44.7^{45.0}_{44.3}$&$45.3^{45.5}_{45.0}$&$45.8^{46.0}_{
45.5}$&$45.5^{45.8}_{45.1}$&$46.1^{46.4}_{45.8}$&$46.8^{47.0}_{46.5}$&$46.3^{46.6}_{46.0}$&$46.9^{47.2}_{46.6}$&$47.5^{47.8}_{
47.2}$\\
16.25&$43.7^{44.0}_{43.3}$&$44.3^{44.5}_{44.0}$&$44.8^{45.0}_{44.5}$&$44.5^{44.8}_{44.2}$&$45.2^{45.5}_{44.8}$&$45.8^{46.0}_{
45.5}$&$45.4^{45.6}_{45.0}$&$45.9^{46.2}_{45.7}$&$46.6^{46.9}_{46.2}$&$46.2^{46.5}_{45.9}$&$46.8^{47.0}_{46.5}$&$47.3^{47.6}_{
47.1}$\\
16.50&$43.5^{43.8}_{43.2}$&$44.2^{44.5}_{43.9}$&$44.8^{45.0}_{44.5}$&$44.4^{44.7}_{44.1}$&$45.0^{45.2}_{44.7}$&$45.6^{45.9}_{
45.3}$&$45.3^{45.5}_{44.9}$&$45.8^{46.1}_{45.5}$&$46.4^{46.6}_{46.1}$&$46.1^{46.4}_{45.8}$&$46.7^{46.9}_{46.4}$&$47.2^{47.5}_{
46.9}$\\
16.75&$43.4^{43.7}_{43.1}$&$44.0^{44.3}_{43.7}$&$44.6^{44.9}_{44.3}$&$44.3^{44.5}_{44.0}$&$44.8^{45.1}_{44.6}$&$45.4^{45.7}_{
45.1}$&$45.1^{45.4}_{44.8}$&$45.7^{45.9}_{45.4}$&$46.2^{46.5}_{46.0}$&$46.0^{46.3}_{45.8}$&$46.5^{46.8}_{46.3}$&$47.1^{47.3}_{
46.8}$\\
17.00&$43.3^{43.5}_{43.0}$&$43.8^{44.1}_{43.6}$&$44.4^{44.7}_{44.1}$&$44.1^{44.4}_{43.8}$&$44.7^{44.9}_{44.4}$&$45.2^{45.5}_{
44.9}$&$45.0^{45.2}_{44.7}$&$45.5^{45.8}_{45.3}$&$46.1^{46.3}_{45.8}$&$45.9^{46.1}_{45.7}$&$46.4^{46.6}_{46.1}$&$46.9^{47.1}_{
46.6}$\\
17.25&$43.1^{43.3}_{42.8}$&$43.6^{43.8}_{43.3}$&$44.2^{44.4}_{43.9}$&$43.9^{44.2}_{43.7}$&$44.4^{44.7}_{44.2}$&$45.0^{45.2}_{
44.7}$&$44.9^{45.0}_{44.6}$&$45.3^{45.5}_{45.1}$&$45.8^{46.0}_{45.5}$&$45.8^{45.9}_{45.6}$&$46.2^{46.4}_{46.0}$&$46.6^{46.9}_{
46.4}$\\
17.50&$42.8^{43.0}_{42.6}$&$43.2^{43.4}_{43.0}$&$43.7^{43.9}_{43.4}$&$43.7^{43.9}_{43.6}$&$44.1^{44.2}_{43.9}$&$44.5^{44.7}_{
44.3}$&$44.7^{44.8}_{44.5}$&$45.0^{45.1}_{44.8}$&$45.3^{45.5}_{45.1}$&$45.6^{45.7}_{45.5}$&$45.9^{46.0}_{45.7}$&$46.2^{46.4}_{
46.0}$\\
17.75&$42.7^{42.8}_{42.6}$&$42.9^{43.0}_{42.8}$&$43.1^{43.3}_{43.0}$&$43.6^{43.7}_{43.5}$&$43.8^{43.9}_{43.7}$&$44.0^{44.1}_{
43.9}$&$44.6^{44.7}_{44.5}$&$44.7^{44.8}_{44.7}$&$44.9^{45.0}_{44.8}$&$45.6^{45.6}_{45.5}$&$45.7^{45.7}_{45.6}$&$45.8^{45.9}_{
45.7}$\\
18.00&$42.7^{42.7}_{42.6}$&$42.8^{42.9}_{42.7}$&$43.0^{43.1}_{42.9}$&$43.6^{43.7}_{43.6}$&$43.7^{43.8}_{43.7}$&$43.9^{44.0}_{
43.8}$&$44.6^{44.6}_{44.6}$&$44.7^{44.7}_{44.6}$&$44.8^{44.8}_{44.7}$&$45.6^{45.6}_{45.5}$&$45.6^{45.6}_{45.6}$&$45.7^{45.7}_{
45.6}$\\
18.25&$42.7^{42.7}_{42.6}$&$42.8^{42.8}_{42.7}$&$42.9^{43.0}_{42.8}$&$43.6^{43.7}_{43.6}$&$43.7^{43.7}_{43.7}$&$43.8^{43.8}_{
43.7}$&$44.6^{44.6}_{44.6}$&$44.6^{44.6}_{44.6}$&$44.7^{44.7}_{44.6}$&$45.6^{45.6}_{45.6}$&$45.6^{45.6}_{45.6}$&$45.5^{45.6}_{
45.5}$\\
18.50&$42.7^{42.7}_{42.6}$&$42.7^{42.8}_{42.7}$&$42.8^{42.9}_{42.8}$&$43.6^{43.7}_{43.6}$&$43.7^{43.7}_{43.7}$&$43.7^{43.8}_{
43.7}$&$44.6^{44.6}_{44.6}$&$44.6^{44.6}_{44.6}$&$44.6^{44.6}_{44.6}$&$45.6^{45.6}_{45.6}$&$45.5^{45.6}_{45.5}$&$45.5^{45.5}_{
45.5}$\\[1ex]

\hline
    \end{tabular}
    \label{tab:fig7}
\end{sidewaystable}


\begin{sidewaystable}
\renewcommand{\arraystretch}{1.2}
\tablenum{A4}
\caption{Bolometric correction
as a function of Eddington ratio (Figure \ref{fig:bc_mdot})}
\tiny
\hspace{-0.7in}
    \begin{tabular}{c@{\hspace{2em}}ccc@{\hspace{3em}}ccc@{\hspace{3em}}ccc@{\hspace{3em}}ccc}
         \hline\hline
         \textbf{spin = 0}
         &\multicolumn{3}{c}{$M_{\rm BH}/M_{\odot} = 10^7$}
         &\multicolumn{3}{c}{$M_{\rm BH}/M_{\odot} = 10^8$}
         &\multicolumn{3}{c}{$M_{\rm BH}/M_{\odot} = 10^9$}
         &\multicolumn{3}{c}{$M_{\rm BH}/M_{\odot} = 10^{10}$}\\[0.5ex]
          \hline 
          $\lambda_{\rm Edd}=$
          &$-1.5$--$-1.0$
          &$-1.0$--$-0.5$
          &$-0.5$--$0.0$
          &$-1.5$--$-1.0$
          &$-1.0$--$-0.5$
          &$-0.5$--$0.0$
          &$-1.5$--$-1.0$
          &$-1.0$--$-0.5$
          &$-0.5$--$0.0$
          &$-1.5$--$-1.0$
          &$-1.0$--$-0.5$
          &$-0.5$--$0.0$\\[0.5ex]
          \hline
$\log(\nu)$\\
14.00&$67.82^{85.75}_{42.72}$&$106.1^{123.3}_{87.66}$&$146.5^{168.1}_{125.3}$&$27.23^{34.46}_{17.60}$&$42.79^{49.92}_{35.22}$&
$59.65^{68.82}_{50.76}$&$11.53^{14.46}_{8.096}$&$17.96^{21.01}_{14.78}$&$25.23^{29.25}_{21.37}$&$5.331^{6.452}_{4.616}$&$7.917
^{9.249}_{6.582}$&$11.13^{12.95}_{9.407}$\\
14.25&$67.82^{85.75}_{42.72}$&$106.1^{123.3}_{87.66}$&$146.5^{168.1}_{125.3}$&$27.23^{34.46}_{17.60}$&$42.79^{49.92}_{35.22}$&
$59.65^{68.82}_{50.76}$&$11.53^{14.46}_{8.096}$&$17.96^{21.01}_{14.78}$&$25.23^{29.25}_{21.37}$&$5.331^{6.452}_{4.616}$&$7.917
^{9.249}_{6.582}$&$11.13^{12.95}_{9.407}$\\
14.50&$41.58^{52.95}_{25.95}$&$66.06^{77.30}_{54.17}$&$92.64^{107.1}_{78.61}$&$17.44^{22.19}_{11.26}$&$27.77^{32.61}_{22.70}$&
$39.29^{45.66}_{33.18}$&$7.846^{9.830}_{5.628}$&$12.25^{14.40}_{10.05}$&$17.40^{20.29}_{14.65}$&$3.978^{4.727}_{3.593}$&$5.758
^{6.716}_{4.817}$&$8.091^{9.431}_{6.831}$\\
14.75&$13.97^{18.11}_{8.596}$&$23.10^{27.51}_{18.57}$&$33.70^{39.66}_{28.03}$&$6.836^{8.697}_{4.615}$&$10.99^{13.05}_{8.901}$&
$15.95^{18.75}_{13.29}$&$3.826^{4.584}_{3.258}$&$5.602^{6.548}_{4.672}$&$7.905^{9.229}_{6.661}$&$2.809^{3.556}_{2.746}$&$3.198
^{3.588}_{2.876}$&$4.191^{4.800}_{3.637}$\\
15.00&$6.665^{8.535}_{4.401}$&$10.85^{12.94}_{8.744}$&$15.88^{18.72}_{13.18}$&$3.921^{4.686}_{3.314}$&$5.717^{6.674}_{4.776}$&
$8.047^{9.384}_{6.789}$&$3.037^{3.876}_{2.976}$&$3.386^{3.771}_{3.078}$&$4.372^{4.982}_{3.820}$&$3.655^{5.457}_{2.794}$&$2.568
^{2.751}_{2.542}$&$2.737^{2.965}_{2.579}$\\
15.25&$4.011^{4.774}_{3.438}$&$5.809^{6.774}_{4.864}$&$8.156^{9.501}_{6.889}$&$3.168^{4.191}_{3.102}$&$3.488^{3.871}_{3.188}$&
$4.472^{5.083}_{3.919}$&$3.994^{6.312}_{2.990}$&$2.708^{2.941}_{2.679}$&$2.850^{3.073}_{2.704}$&$6.186^{7.048}_{4.595}$&$3.403
^{4.442}_{2.783}$&$2.497^{2.744}_{2.398}$\\
15.50&$3.214^{4.320}_{3.151}$&$3.536^{3.920}_{3.235}$&$4.525^{5.139}_{3.969}$&$4.120^{6.743}_{3.069}$&$2.773^{3.019}_{2.739}$&
$2.905^{3.126}_{2.761}$&$6.660^{8.101}_{4.838}$&$3.556^{4.674}_{2.892}$&$2.582^{2.850}_{2.472}$&$7.953^{8.804}_{7.496}$&$6.674
^{7.444}_{5.315}$&$4.079^{5.178}_{3.241}$\\
15.75&$4.177^{6.878}_{3.097}$&$2.785^{3.042}_{2.744}$&$2.896^{3.110}_{2.761}$&$6.820^{8.634}_{4.963}$&$3.648^{4.796}_{2.952}$&
$2.622^{2.905}_{2.495}$&$8.571^{10.12}_{7.874}$&$6.971^{7.810}_{5.556}$&$4.253^{5.414}_{3.367}$&$9.961^{10.91}_{9.499}$&$9.499
^{9.563}_{9.459}$&$9.140^{9.542}_{7.881}$\\
16.00&$6.651^{8.639}_{4.845}$&$3.593^{4.683}_{2.926}$&$2.611^{2.883}_{2.492}$&$8.645^{10.65}_{7.813}$&$6.865^{7.740}_{5.475}$&
$4.212^{5.332}_{3.343}$&$10.65^{12.36}_{9.899}$&$9.725^{9.864}_{9.664}$&$9.149^{9.639}_{7.838}$&$12.32^{13.02}_{11.99}$&$12.11
^{12.40}_{11.99}$&$12.78^{13.03}_{12.43}$\\
16.25&$8.413^{10.62}_{7.514}$&$6.573^{7.437}_{5.266}$&$4.089^{5.136}_{3.271}$&$10.75^{12.94}_{9.857}$&$9.579^{9.811}_{9.426}$&
$8.857^{9.401}_{7.578}$&$13.18^{14.64}_{12.55}$&$12.50^{12.63}_{12.45}$&$12.85^{12.95}_{12.64}$&$14.98^{15.07}_{14.93}$&$15.63
^{16.22}_{15.11}$&$16.92^{17.40}_{16.28}$\\
16.50&$10.54^{12.89}_{9.571}$&$9.213^{9.518}_{8.981}$&$8.375^{8.944}_{7.164}$&$13.38^{15.25}_{12.62}$&$12.45^{12.58}_{12.43}$&
$12.56^{12.60}_{12.49}$&$16.02^{16.69}_{15.88}$&$16.28^{16.73}_{15.93}$&$17.23^{17.51}_{16.77}$&$17.81^{18.79}_{16.50}$&$20.22
^{21.43}_{18.92}$&$22.74^{23.63}_{21.55}$\\
16.75&$13.42^{15.30}_{12.65}$&$12.39^{12.61}_{12.33}$&$12.31^{12.34}_{12.18}$&$16.57^{17.41}_{16.43}$&$16.79^{17.16}_{16.48}$&
$17.53^{17.70}_{17.21}$&$19.34^{20.33}_{18.31}$&$21.74^{22.90}_{20.45}$&$24.07^{24.74}_{23.00}$&$20.92^{23.48}_{17.43}$&$26.64
^{29.09}_{23.78}$&$31.66^{33.37}_{29.33}$\\
17.00&$17.21^{17.62}_{17.13}$&$17.67^{18.03}_{17.30}$&$18.28^{18.34}_{18.06}$&$20.51^{21.85}_{19.15}$&$23.55^{24.82}_{22.01}$&
$26.02^{26.65}_{24.97}$&$23.13^{26.23}_{19.31}$&$29.95^{32.75}_{26.58}$&$35.52^{37.15}_{33.00}$&$24.15^{29.43}_{17.69}$&$35.99
^{41.09}_{30.06}$&$46.44^{49.99}_{41.59}$\\
17.25&$23.15^{25.92}_{19.99}$&$29.08^{31.31}_{26.25}$&$33.26^{34.16}_{31.51}$&$26.16^{31.21}_{20.56}$&$37.38^{41.99}_{31.79}$&
$46.55^{49.27}_{42.51}$&$28.00^{35.73}_{19.78}$&$45.83^{53.99}_{36.64}$&$62.61^{68.08}_{54.77}$&$27.91^{38.31}_{17.45}$&$53.13
^{65.99}_{39.64}$&$80.68^{91.04}_{67.33}$\\
17.50&$31.11^{42.01}_{21.23}$&$57.98^{72.53}_{43.43}$&$89.67^{101.7}_{74.10}$&$32.56^{46.36}_{20.74}$&$68.44^{90.22}_{48.14}$&
$118.7^{141.1}_{92.98}$&$32.76^{49.33}_{19.19}$&$77.84^{108.5}_{51.55}$&$152.0^{188.6}_{112.0}$&$31.12^{49.73}_{16.42}$&$84.38
^{124.7}_{52.40}$&$187.2^{246.1}_{129.6}$\\
17.75&$32.89^{51.15}_{19.19}$&$85.72^{128.7}_{53.81}$&$203.4^{285.7}_{134.2}$&$33.32^{53.81}_{18.45}$&$94.75^{148.1}_{56.72}$&
$247.8^{368.1}_{156.0}$&$32.75^{55.26}_{16.88}$&$102.5^{167.8}_{58.54}$&$297.6^{462.9}_{176.3}$&$30.58^{54.21}_{14.31}$&$106.7
^{183.4}_{57.86}$&$347.6^{581.2}_{194.1}$\\
18.00&$30.80^{51.14}_{16.41}$&$92.42^{147.8}_{54.20}$&$253.7^{383.8}_{155.3}$&$31.15^{53.70}_{15.77}$&$101.9^{169.3}_{56.99}$&
$307.0^{491.1}_{179.7}$&$30.59^{55.07}_{14.41}$&$109.9^{191.3}_{58.75}$&$367.1^{613.7}_{202.3}$&$28.55^{53.97}_{12.21}$&$114.2
^{208.5}_{58.01}$&$427.2^{768.9}_{222.1}$\\
18.25&$28.71^{50.83}_{13.96}$&$98.62^{167.0}_{54.25}$&$307.4^{493.6}_{176.5}$&$29.03^{53.37}_{13.41}$&$108.7^{191.2}_{57.03}$&
$371.8^{631.6}_{204.3}$&$28.51^{54.74}_{12.26}$&$117.3^{216.1}_{58.78}$&$444.5^{788.1}_{229.9}$&$26.61^{53.63}_{10.38}$&$121.9
^{235.4}_{58.05}$&$517.1^{988.9}_{252.3}$\\
18.50&$27.70^{50.75}_{12.83}$&$102.3^{178.5}_{54.37}$&$341.0^{565.0}_{189.3}$&$28.01^{53.28}_{12.32}$&$112.7^{204.3}_{57.14}$&
$412.3^{723.0}_{219.1}$&$27.50^{54.65}_{11.26}$&$121.6^{230.9}_{58.90}$&$492.9^{901.6}_{246.4}$&$25.67^{53.54}_{9.536}$&$126.4
^{251.6}_{58.17}$&$573.3^{1132.}_{270.5}$\\[0.5ex]
         \hline
         \textbf{spin = 1}\\
14.00&$206.2^{245.7}_{150.8}$&$294.3^{339.1}_{250.0}$&$403.9^{467.1}_{344.5}$&$86.02^{104.1}_{60.93}$&$126.1^{146.4}_{106.0}$&
$175.8^{204.7}_{148.9}$&$37.25^{45.79}_{25.56}$&$56.16^{65.65}_{46.71}$&$79.35^{92.77}_{66.79}$&$16.79^{20.94}_{11.22}$&$25.93
^{30.47}_{21.39}$&$36.98^{43.38}_{31.01}$\\
14.25&$206.2^{245.7}_{150.8}$&$294.3^{339.1}_{250.0}$&$403.9^{467.1}_{344.5}$&$86.02^{104.1}_{60.93}$&$126.1^{146.4}_{106.0}$&
$175.8^{204.7}_{148.9}$&$37.25^{45.79}_{25.56}$&$56.16^{65.65}_{46.71}$&$79.35^{92.77}_{66.79}$&$16.79^{20.94}_{11.22}$&$25.93
^{30.47}_{21.39}$&$36.98^{43.38}_{31.01}$\\
14.50&$133.8^{162.1}_{94.79}$&$197.0^{229.2}_{165.2}$&$275.8^{321.6}_{233.0}$&$58.14^{71.48}_{39.83}$&$87.83^{102.8}_{72.95}$&
$124.6^{145.9}_{104.6}$&$26.26^{32.73}_{17.50}$&$40.57^{47.73}_{33.43}$&$58.03^{68.13}_{48.58}$&$12.37^{15.55}_{8.127}$&$19.37
^{22.81}_{15.89}$&$27.75^{32.59}_{23.23}$\\
14.75&$51.78^{65.06}_{33.81}$&$81.31^{96.20}_{66.51}$&$117.7^{138.8}_{97.99}$&$24.66^{31.14}_{15.85}$&$39.00^{46.12}_{31.86}$&
$56.35^{66.38}_{46.97}$&$12.13^{15.31}_{7.863}$&$19.12^{22.54}_{15.65}$&$27.44^{32.20}_{22.95}$&$6.299^{7.831}_{4.303}$&$9.666
^{11.31}_{7.997}$&$13.66^{15.94}_{11.51}$\\
15.00&$24.62^{31.21}_{15.68}$&$39.16^{46.36}_{31.92}$&$56.69^{66.78}_{47.22}$&$12.27^{15.48}_{7.953}$&$19.32^{22.77}_{15.83}$&
$27.70^{32.51}_{23.18}$&$6.460^{7.999}_{4.466}$&$9.848^{11.51}_{8.165}$&$13.87^{16.16}_{11.70}$&$3.832^{4.501}_{3.113}$&$5.347
^{6.123}_{4.576}$&$7.242^{8.339}_{6.217}$\\
15.25&$12.34^{15.57}_{8.024}$&$19.44^{22.92}_{15.92}$&$27.89^{32.72}_{23.34}$&$6.537^{8.083}_{4.539}$&$9.949^{11.62}_{8.254}$&
$14.00^{16.31}_{11.82}$&$3.919^{4.587}_{3.220}$&$5.439^{6.221}_{4.662}$&$7.346^{8.447}_{6.314}$&$3.101^{3.481}_{3.057}$&$3.416
^{3.722}_{3.168}$&$4.208^{4.708}_{3.761}$\\
15.50&$6.560^{8.119}_{4.564}$&$9.999^{11.68}_{8.289}$&$14.08^{16.41}_{11.88}$&$3.948^{4.621}_{3.251}$&$5.481^{6.270}_{4.699}$&
$7.403^{8.514}_{6.364}$&$3.149^{3.553}_{3.101}$&$3.460^{3.768}_{3.210}$&$4.257^{4.758}_{3.807}$&$3.974^{5.891}_{3.213}$&$2.935
^{3.170}_{2.871}$&$2.947^{3.103}_{2.872}$\\
15.75&$3.890^{4.550}_{3.222}$&$5.396^{6.172}_{4.625}$&$7.289^{8.380}_{6.266}$&$3.136^{3.573}_{3.087}$&$3.428^{3.729}_{3.186}$&
$4.208^{4.702}_{3.768}$&$4.019^{5.977}_{3.240}$&$2.948^{3.196}_{2.873}$&$2.942^{3.087}_{2.873}$&$6.648^{8.251}_{5.220}$&$4.152
^{5.093}_{3.454}$&$3.030^{3.401}_{2.803}$\\
16.00&$3.088^{3.484}_{3.043}$&$3.409^{3.718}_{3.157}$&$4.204^{4.702}_{3.757}$&$3.912^{5.785}_{3.180}$&$2.914^{3.137}_{2.849}$&
$2.927^{3.080}_{2.852}$&$6.480^{8.174}_{5.086}$&$4.063^{4.965}_{3.396}$&$2.992^{3.344}_{2.776}$&$8.520^{10.26}_{7.556}$&$6.824
^{7.480}_{5.984}$&$5.043^{5.892}_{4.187}$\\
16.25&$3.739^{5.458}_{3.076}$&$2.842^{3.038}_{2.797}$&$2.889^{3.047}_{2.801}$&$6.147^{7.851}_{4.833}$&$3.897^{4.717}_{3.281}$&
$2.917^{3.237}_{2.722}$&$8.307^{10.18}_{7.296}$&$6.541^{7.216}_{5.719}$&$4.829^{5.630}_{4.027}$&$10.74^{12.81}_{9.540}$&$8.844
^{9.454}_{8.349}$&$7.879^{8.304}_{7.275}$\\
16.50&$5.697^{7.375}_{4.485}$&$3.651^{4.383}_{3.116}$&$2.791^{3.072}_{2.625}$&$7.922^{9.850}_{6.881}$&$6.121^{6.797}_{5.335}$&
$4.526^{5.255}_{3.803}$&$10.57^{12.78}_{9.317}$&$8.566^{9.226}_{8.017}$&$7.498^{7.968}_{6.871}$&$13.64^{15.76}_{12.38}$&$11.61
^{12.29}_{11.01}$&$10.47^{10.95}_{9.915}$\\
16.75&$7.542^{9.525}_{6.445}$&$5.669^{6.359}_{4.930}$&$4.194^{4.854}_{3.549}$&$10.41^{12.69}_{9.104}$&$8.290^{9.006}_{7.674}$&
$7.086^{7.613}_{6.436}$&$13.83^{16.06}_{12.51}$&$11.67^{12.41}_{11.00}$&$10.37^{10.94}_{9.737}$&$17.49^{19.23}_{16.48}$&$15.86
^{16.41}_{15.32}$&$14.74^{15.26}_{14.05}$\\
17.00&$10.47^{12.82}_{9.063}$&$8.139^{8.955}_{7.399}$&$6.710^{7.334}_{5.976}$&$14.28^{16.51}_{12.93}$&$12.00^{12.82}_{11.23}$&
$10.46^{11.15}_{9.682}$&$18.44^{20.13}_{17.49}$&$16.87^{17.42}_{16.26}$&$15.57^{16.20}_{14.74}$&$22.63^{23.21}_{22.55}$&$22.62
^{22.66}_{22.52}$&$22.12^{22.49}_{21.37}$\\
17.25&$16.74^{18.74}_{15.44}$&$14.42^{15.33}_{13.41}$&$12.37^{13.32}_{11.25}$&$21.66^{22.65}_{21.23}$&$20.87^{21.22}_{20.35}$&
$19.43^{20.25}_{18.30}$&$26.62^{27.57}_{26.04}$&$28.58^{28.96}_{27.68}$&$28.77^{28.99}_{27.93}$&$31.24^{34.17}_{28.42}$&$37.54
^{39.39}_{34.51}$&$40.45^{40.65}_{39.55}$\\
17.50&$31.18^{34.10}_{28.47}$&$37.25^{38.55}_{34.45}$&$38.78^{39.01}_{37.46}$&$36.26^{42.01}_{31.08}$&$49.47^{54.55}_{42.79}$&
$58.16^{59.26}_{54.81}$&$40.61^{49.61}_{32.92}$&$62.33^{72.71}_{50.74}$&$82.32^{88.24}_{73.94}$&$44.00^{56.69}_{33.63}$&$76.40
^{93.35}_{58.37}$&$113.0^{124.3}_{95.23}$\\
17.75&$41.65^{54.57}_{31.78}$&$76.36^{97.11}_{56.38}$&$125.9^{150.9}_{100.6}$&$44.83^{60.81}_{32.98}$&$89.89^{121.7}_{63.27}$&
$167.5^{208.8}_{124.6}$&$47.52^{66.85}_{33.73}$&$103.4^{147.5}_{69.57}$&$217.0^{296.7}_{153.9}$&$49.46^{72.47}_{33.58}$&$119.0
^{176.2}_{75.87}$&$281.9^{393.7}_{184.0}$\\
18.00&$41.49^{57.91}_{29.61}$&$88.14^{120.4}_{60.29}$&$172.4^{227.5}_{126.2}$&$44.52^{64.19}_{30.68}$&$103.0^{149.9}_{67.31}$&
$226.7^{307.9}_{154.4}$&$47.07^{70.32}_{31.34}$&$117.7^{180.3}_{73.70}$&$290.9^{436.9}_{189.8}$&$48.93^{76.09}_{31.17}$&$135.2
^{214.0}_{80.23}$&$376.7^{572.4}_{225.3}$\\
18.25&$40.90^{60.50}_{27.41}$&$99.12^{143.8}_{63.43}$&$222.8^{316.2}_{152.1}$&$43.89^{67.04}_{28.40}$&$115.8^{179.4}_{70.82}$&
$293.2^{425.8}_{185.8}$&$46.40^{73.44}_{29.00}$&$132.2^{215.6}_{77.48}$&$375.7^{607.2}_{228.8}$&$48.23^{79.50}_{28.85}$&$152.0
^{255.7}_{84.38}$&$487.3^{790.9}_{271.0}$\\
18.50&$40.67^{62.03}_{26.35}$&$105.7^{158.2}_{65.27}$&$255.6^{377.2}_{168.2}$&$43.64^{68.73}_{27.29}$&$123.5^{197.7}_{72.89}$&
$336.7^{506.3}_{205.4}$&$46.13^{75.28}_{27.88}$&$140.8^{237.4}_{79.70}$&$431.1^{724.3}_{253.0}$&$47.94^{81.51}_{27.73}$&$162.0
^{281.5}_{86.83}$&$559.6^{940.1}_{299.5}$\\[1ex]
\hline
\end{tabular}
\label{tab:fig8}
\end{sidewaystable}


%% file: main.bbl
\begin{thebibliography}{33}
\expandafter\ifx\csname natexlab\endcsname\relax\def\natexlab#1{#1}\fi

\bibitem[{{Azadi} {et~al.}(2020){Azadi}, {Wilkes}, {Kuraszkiewicz}, {McDowell},
  {Siebenmorgen}, {Ashby}, {Birkinshaw}, {Worrall}, {Abrams}, {Barthel},
  {Fazio}, {Haas}, {Hyman}, {Mart{\'\i}nez-Galarza}, \& {Meyer}}]{Azadi2020}
{Azadi}, M., {et~al.} 2020, \href {http://arxiv.org/abs/2011.03130} {arXiv
  e-prints}, arXiv:2011.03130

\bibitem[{{Blandford} \& {K{\"o}nigl}(1979)}]{Blandford1979}
{Blandford}, R.~D., \& {K{\"o}nigl}, A. 1979, \href
  {http://dx.doi.org/10.1086/157262} {\apj}, 232, 34

\bibitem[{{Brenneman} \& {Reynolds}(2006)}]{lb2006}
{Brenneman}, L.~W., \& {Reynolds}, C.~S. 2006, \href
  {http://dx.doi.org/10.1086/508146} {\apj}, 652, 1028

\bibitem[{Charlot \& Fall(2000)}]{charlot2000simple}
Charlot, S., \& Fall, S. 2000, ApJ, 539, 718

\bibitem[{{da Cunha} {et~al.}(2008){da Cunha}, {Charlot}, \& {Elbaz}}]{dc2008}
{da Cunha}, E., {Charlot}, S., \& {Elbaz}, D. 2008, \href
  {http://dx.doi.org/10.1111/j.1365-2966.2008.13535.x} {\mnras}, 388, 1595

\bibitem[{{da Cunha} {et~al.}(2015){da Cunha}, {Walter}, {Smail}, {Swinbank},
  {Simpson}, {Decarli}, {Hodge}, {Weiss}, {van der Werf}, {Bertoldi},
  {Chapman}, {Cox}, {Danielson}, {Dannerbauer}, {Greve}, {Ivison}, {Karim}, \&
  {Thomson}}]{dcr2015}
{da Cunha}, E., {et~al.} 2015, \href
  {http://dx.doi.org/10.1088/0004-637X/806/1/110} {\apj}, 806, 110

\bibitem[{{Elvis} {et~al.}(1994){Elvis}, {Wilkes}, {McDowell}, {Green},
  {Bechtold}, {Willner}, {Oey}, {Polomski}, \& {Cutri}}]{Elvis1994}
{Elvis}, M., {et~al.} 1994, \href {http://dx.doi.org/10.1086/192093} {\apjs},
  95, 1

\bibitem[{{Frank} {et~al.}(2002){Frank}, {King}, \& {Raine}}]{FRK2002}
{Frank}, J., {King}, A., \& {Raine}, D.~J. 2002, {Accretion Power in
  Astrophysics: Third Edition}

\bibitem[{{Ho}(2008)}]{Ho2008}
{Ho}, L.~C. 2008, \href
  {http://dx.doi.org/10.1146/annurev.astro.45.051806.110546} {\araa}, 46, 475

\bibitem[{{Hopkins} {et~al.}(2007){Hopkins}, {Richards}, \&
  {Hernquist}}]{Hopkins2007}
{Hopkins}, P.~F., {Richards}, G.~T., \& {Hernquist}, L. 2007, \href
  {http://dx.doi.org/10.1086/509629} {\apj}, 654, 731

\bibitem[{{Hubeny} {et~al.}(2001){Hubeny}, {Blaes}, {Krolik}, \&
  {Agol}}]{Hubeny2001}
{Hubeny}, I., {et~al.} 2001, \href {http://dx.doi.org/10.1086/322344} {\apj},
  559, 680

\bibitem[{{Konigl}(1981)}]{Konigl1981}
{Konigl}, A. 1981, \href {http://dx.doi.org/10.1086/158638} {\apj}, 243, 700

\bibitem[{{Korista} {et~al.}(1997){Korista}, {Baldwin}, {Ferland}, \&
  {Verner}}]{K1997}
{Korista}, K., {et~al.} 1997, \href {http://dx.doi.org/10.1086/312966} {\apjs},
  108, 401

\bibitem[{{Kubota} \& {Done}(2018)}]{Kubota2018}
{Kubota}, A., \& {Done}, C. 2018, \href
  {http://dx.doi.org/10.1093/mnras/sty1890} {\mnras}, 480, 1247

\bibitem[{{Laing} {et~al.}(1983){Laing}, {Riley}, \& {Longair}}]{Laing1983}
{Laing}, R.~A., {Riley}, J.~M., \& {Longair}, M.~S. 1983, \href
  {http://dx.doi.org/10.1093/mnras/204.1.151} {\mnras}, 204, 151

\bibitem[{{Leipski} {et~al.}(2010){Leipski}, {Haas}, {Willner}, {Ashby},
  {Wilkes}, {Fazio}, {Antonucci}, {Barthel}, {Chini}, {Siebenmorgen}, {Ogle},
  \& {Heymann}}]{Leipski2010}
{Leipski}, C., {et~al.} 2010, \href
  {http://dx.doi.org/10.1088/0004-637X/717/2/766} {\apj}, 717, 766

\bibitem[{{Marconi} {et~al.}(2004){Marconi}, {Risaliti}, {Gilli}, {Hunt},
  {Maiolino}, \& {Salvati}}]{Marconi2004}
{Marconi}, A., {et~al.} 2004, \href
  {http://dx.doi.org/10.1111/j.1365-2966.2004.07765.x} {\mnras}, 351, 169

\bibitem[{{Mathews} \& {Ferland}(1987)}]{FM1987}
{Mathews}, W.~G., \& {Ferland}, G.~J. 1987, \href
  {http://dx.doi.org/10.1086/165843} {\apj}, 323, 456

\bibitem[{{Nemmen} \& {Brotherton}(2010)}]{Nemmen2010}
{Nemmen}, R.~S., \& {Brotherton}, M.~S. 2010, \href
  {http://dx.doi.org/10.1111/j.1365-2966.2010.17224.x} {\mnras}, 408, 1598

\bibitem[{{Novikov} \& {Thorne}(1973)}]{NT1973}
{Novikov}, I.~D., \& {Thorne}, K.~S. 1973, in Black Holes (Les Astres Occlus),
  343--450

\bibitem[{{Podigachoski} {et~al.}(2015){Podigachoski}, {Barthel}, {Haas},
  {Leipski}, {Wilkes}, {Kuraszkiewicz}, {Westhues}, {Willner}, {Ashby},
  {Chini}, {Clements}, {Fazio}, {Labiano}, {Lawrence}, {Meisenheimer},
  {Peletier}, {Siebenmorgen}, \& {Verdoes Kleijn}}]{pece15}
{Podigachoski}, P., {et~al.} 2015, \href
  {http://dx.doi.org/10.1051/0004-6361/201425137} {\aap}, 575, A80

\bibitem[{{Porquet} {et~al.}(2018){Porquet}, {Reeves}, {Matt}, {Marinucci},
  {Nardini}, {Braito}, {Lobban}, {Ballantyne}, {Boggs}, {Christensen},
  {Dauser}, {Farrah}, {Garcia}, {Hailey}, {Harrison}, {Stern}, {Tortosa},
  {Ursini}, \& {Zhang}}]{Porquet2018}
{Porquet}, D., {et~al.} 2018, \href
  {http://dx.doi.org/10.1051/0004-6361/201731290} {\aap}, 609, A42

\bibitem[{{Richards} {et~al.}(2006){Richards}, {Lacy}, {Storrie-Lombardi},
  {Hall}, {Gallagher}, {Hines}, {Fan}, {Papovich}, {Vanden Berk}, {Trammell},
  {Schneider}, {Vestergaard}, {York}, {Jester}, {Anderson}, {Budav{\'a}ri}, \&
  {Szalay}}]{Richards2006}
{Richards}, G.~T., {et~al.} 2006, \href {http://dx.doi.org/10.1086/506525}
  {\apjs}, 166, 470

\bibitem[{{Runnoe} {et~al.}(2012){Runnoe}, {Brotherton}, \& {Shang}}]{R12}
{Runnoe}, J.~C., {Brotherton}, M.~S., \& {Shang}, Z. 2012, \href
  {http://dx.doi.org/10.1111/j.1365-2966.2012.20620.x} {\mnras}, 422, 478

\bibitem[{{Shakura} \& {Sunyaev}(1976)}]{SS1976}
{Shakura}, N.~I., \& {Sunyaev}, R.~A. 1976, \href
  {http://dx.doi.org/10.1093/mnras/175.3.613} {\mnras}, 175, 613

\bibitem[{{Shang} {et~al.}(2011){Shang}, {Brotherton}, {Wills}, {Wills},
  {Cales}, {Dale}, {Green}, {Runnoe}, {Nemmen}, {Gallagher}, {Ganguly},
  {Hines}, {Kelly}, {Kriss}, {Li}, {Tang}, \& {Xie}}]{Shang2011}
{Shang}, Z., {et~al.} 2011, \href {http://dx.doi.org/10.1088/0067-0049/196/1/2}
  {\apjs}, 196, 2

\bibitem[{{Siebenmorgen} {et~al.}(2015){Siebenmorgen}, {Heymann}, \&
  {Efstathiou}}]{Siebenmorgen2015}
{Siebenmorgen}, R., {Heymann}, F., \& {Efstathiou}, A. 2015, \href
  {http://dx.doi.org/10.1051/0004-6361/201526034} {\aap}, 583, A120

\bibitem[{{Vasudevan} \& {Fabian}(2007)}]{V2007}
{Vasudevan}, R.~V., \& {Fabian}, A.~C. 2007, \href
  {http://dx.doi.org/10.1111/j.1365-2966.2007.12328.x} {\mnras}, 381, 1235

\bibitem[{{Vignali} {et~al.}(2003){Vignali}, {Brandt}, \& {Schneider}}]{V2003}
{Vignali}, C., {Brandt}, W.~N., \& {Schneider}, D.~P. 2003, \href
  {http://dx.doi.org/10.1086/345973} {\aj}, 125, 433

\bibitem[{{Wilkes} \& {Elvis}(1987)}]{Wilkes1987}
{Wilkes}, B.~J., \& {Elvis}, M. 1987, \href {http://dx.doi.org/10.1086/165822}
  {\apj}, 323, 243

\bibitem[{{Wilkes} {et~al.}(2013){Wilkes}, {Kuraszkiewicz}, {Haas}, {Barthel},
  {Leipski}, {Willner}, {Worrall}, {Birkinshaw}, {Antonucci}, {Ashby}, {Chini},
  {Fazio}, {Lawrence}, {Ogle}, \& {Schulz}}]{Wilkes2013}
{Wilkes}, B.~J., {et~al.} 2013, \href
  {http://dx.doi.org/10.1088/0004-637X/773/1/15} {\apj}, 773, 15

\bibitem[{{Wright}(2006)}]{2006PASP..118.1711W}
{Wright}, E.~L. 2006, \href {http://dx.doi.org/10.1086/510102} {\pasp}, 118,
  1711

\bibitem[{{Yaqoob} {et~al.}(2016){Yaqoob}, {Turner}, {Tatum}, {Trevor}, \&
  {Scholtes}}]{Yaqoob2016}
{Yaqoob}, T., {et~al.} 2016, \href {http://dx.doi.org/10.1093/mnras/stw1824}
  {\mnras}, 462, 4038

\end{thebibliography}
